\documentclass[]{aa}  
\usepackage{amssymb}
\usepackage{graphicx}
\usepackage[authoryear]{natbib}
\usepackage[figuresright]{rotating}
\usepackage{subfigure}

\usepackage{txfonts}
\usepackage{color}
\usepackage[table]{xcolor} 
\usepackage{bm}

\newcommand{\teff}{\ifmmode T_{\rm eff} \else T$_{\mathrm{eff}}$\fi}
\newcommand{\logg}{\ifmmode \log g \else $\log g$\fi}
\newcommand{\lL}{\ifmmode \log \frac{L}{L_{\odot}} \else $\log \frac{L}{L_{\odot}}$\fi}
\newcommand{\mdot}{$\dot{M}$}

\newcommand{\vsini}{$V$ sin$i$}

\newcommand{\kms}{km s$^{-1}$}
\newcommand{\msun}{\ifmmode M_{\odot} \else M$_{\odot}$\fi}
\newcommand{\zsun}{\ifmmode Z_{\odot} \else Z$_{\odot}$\fi}
\newcommand{\lsun}{\ifmmode L_{\odot} \else L$_{\odot}$\fi}
\newcommand{\rsun}{\ifmmode R_{\odot} \else R$_{\odot}$\fi}
\newcommand{\qh}{\ifmmode Q_{\rm H} \else $Q_{\rm H}$\fi}
\newcommand{\qhei}{\ifmmode Q_{\ion{He}{i}} \else $Q_{\ion{He}{i}}$\fi}

\newcommand{\mum}{\ifmmode \mu m \else $\mu m$\fi}

\definecolor{Gray}{gray}{0.9}
\begin{document}
   \title{A comparison of evolutionary tracks for single Galactic massive stars}

   \subtitle{}

   \author{F. Martins\inst{1}
          \and
          A. Palacios\inst{1}
          }

   \offprints{F. Martins}

   \institute{LUPM, Universit\'e Montpellier 2, CNRS, Place Eug\`ene Bataillon, F-34095 Montpellier, France \\
              \email{fabrice.martins AT univ-montp2.fr}
             }

   \date{}

\authorrunning{Martins \& Palacios}
\titlerunning{Comparing evolutionary tracks for Galactic massive stars}

 
  \abstract
   {The evolution of massive stars is not fully understood. The relation between different types of evolved massive stars is not clear, and the role of factors such as binarity, rotation or magnetism needs to be quantified.}
   {Several groups make available the results of 1-D single stellar evolution calculations in the form of evolutionary tracks and isochrones. They use different stellar evolution codes for which the input physics and its implementation varies.  
   In this paper, we aim at comparing the 
   currently available evolutionary tracks for massive stars. We focus on calculations aiming at reproducing the evolution of Galactic stars. Our main goal is to highlight the uncertainties on the predicted evolutionary paths. }
   {We compute stellar evolution models with the codes MESA and STAREVOL. We compare our results with those of four published grids of massive stellar evolution models (Geneva, STERN, Padova and FRANEC codes). We first investigate the effects of overshooting, mass loss, metallicity, chemical composition. We subsequently focus on rotation. Finally, we compare the predictions of published evolutionary models with the observed properties of a large sample of Galactic stars.}
   {We find that all models agree well for the main sequence evolution. Large differences in luminosity and temperatures appear for the post main sequence evolution, especially in the cool part of the Hertzsprung-Russell (HR) diagram. Depending on the physical ingredients, tracks of different initial masses can overlap, rendering any mass estimate doubtful. For masses between 7 and 20 \msun, we find that the main sequence width is slightly too narrow in the Geneva models including rotation. It is (much) too wide for the (STERN) FRANEC models. This conclusion is reached from the investigation of the HR diagram and from the evolution of the surface velocity as a function of surface gravity. An overshooting parameter $\alpha$ between 0.1 and 0.2 in models with rotation is preferred to reproduce the main sequence width. Determinations of surface abundances of carbon and nitrogen are partly inconsistent and cannot be used at present to discriminate between the predictions of published tracks. For stars with initial masses larger than about 60 \msun, the FRANEC models with rotation can reproduce the observations of luminous O supergiants and WNh stars, while the Geneva models remain too hot.}
   {}


   \maketitle


\section{Introduction}
\label{s_intro}

Massive stars (M$>$ 8 \msun) are born as O and B stars on the zero age main sequence (hereafter ZAMS). After a few million years, they evolve off the main sequence to become either (red) supergiants or Wolf-Rayet stars, depending on their initial mass. They may go through a phase during which they are seen as Luminous Blue Variables (LBV), blue or yellow supergiants. Beyond this qualitative scenario, little is known about the evolution of massive stars. In particular, the detailed relations between stars of different types is poorly constrained. Very massive and luminous H-rich WN stars are probably core-H burning objects \citep{arches,paul10}. They may become LBVs when they evolve towards the red part of the Hertzsprung-Russell (HR) diagram \citep{sc08}. At lower masses (M$\sim$50 \msun), \citet{cb97} and \citet{martins07} provided relations between mid-O supergiants and several types of WN and WC stars. These evolutionary sequences remain partial, and do not exist for the entire upper HR diagram. 

Stellar evolutionary models have been developed to explain and predict the physical properties of massive stars. The effective temperature and luminosity they predict are used to build tracks followed by stars of different initial masses in the HR diagram. These predictions are confronted to observations to test the input physics. \citet{hamann06} studied the WN stars in the Galaxy and concluded that the models of \citet{mm03} are able to account for the global properties of WN stars. However, some quantitative problems exist, especially regarding the number of early and late WN stars. Similarly, the ratio of WC to WN stars provides a test of evolutionary models. According to the classical scenario, WC stars represent a more advanced state of evolution than WN stars, simply because of mass loss: as evolution proceeds, mass is removed by stellar winds and deeper layers are unveiled. It takes more time to reach deeper layers, and consequently these layers bear the imprint of the nucleosynthesis occurring at
later stages of evolution. \citet{nm11} showed that the ratio WC/WN is
correctly reproduced by the models of \citet{mm03} at low
metallicity. This is not the case at higher metallicity \citep[see
  also][]{nm12} where evolutionary models predict too few WC
stars. \citet{hunter08} studied the nitrogen content of B stars in the
Large Magellanic Cloud and found that single standard stellar
evolutionary models could account for the properties of roughly two
thirds of the sample, the remaining objects being unexplained by current single star tracks with rotation.
The results were confirmed by the calculations of \citet{brott11b}. Nonetheless, \citet{maeder09} cautioned that many parameters could affect the surface abundances and that they needed to be considered.

Indeed, evolutionary calculations rely on various prescriptions to describe the physical processes driving the evolution, and these prescriptions may vary from code to code. The most important ingredients/processes to be considered are convection and its related properties (such as overshooting), mass loss \citep{cm86}, chemical composition (and the relative abundance of the various species considered in the models) and of course initial mass. Rotation is another key ingredient, since it affects the internal structure, the physical properties (temperature, luminosity), the surface chemical appearance and the lifetimes of stars \citep{mm00}. Several prescriptions usually exist to treat a given physical process in evolutionary codes. As a consequence, the outputs depend on the input physics. Since evolutionary calculations are a crucial tool to link the observed properties of stars to their physical state and evolution, it is important to understand the limitations and uncertainties associated with evolutionary models. 

In this paper, we present a comparison of various predictions of the evolution of massive stars computed with different codes. Our goal is to highlight the uncertainties in the outputs of evolutionary calculations, especially concerning the HR diagram. We focus on calculations for Galactic stars. In Sect.\ \ref{s_models} we describe the different codes and models we have used in our comparisons. We then present a study of the uncertainties associated with the assumptions in the input physics (Sect.\ \ref{s_comp_theo}). In the same section, we also compare the evolutionary tracks predicted by the different codes. In Sect.\ \ref{s_comp_obs} we confront the predictions of published grids of models to the observed properties of massive stars in the Galaxy. We highlight the limitations of each grid. Finally, we summarize our main conclusions in Sect.~\ref{s_conc}.


\section{Stellar evolution models}
\label{s_models}

To achieve our goal of comparing stellar evolution tracks of massive
stars, we have used four databases of stellar evolution tracks
presented in \citet{bert09}, \citet{brott11a}, \citet{ek12},
\citet{cl13} and we computed models using the STAREVOL code
\citep{dmp09} and the MESA code \citep{pax11,pax13}. We recall here the main
ingredients and physical parameters used in each case since they may
differ largely and these differences appear to affect the evolutionary
tracks. Table \ref{tab_mod} summarizes the main inputs for each code.

\begin{table*}[t]
\begin{center}
\caption{Main ingredients of the evolutionary models} \label{tab_mod}
\begin{tabular}{lcccccc}
\hline \hline
 &   STERN$^1$  & Geneva$^2$   & FRANEC$^3$  &   Padova$^4$& MESA$^5$  &   STAREVOL$^5$  \\ 
\hline
Initial metallicity (Z)  &     0.0088   &   0.0140  &  0.01345 &  0.0170              &  0.014   & 0.0134\\
Mixing length parameter ($l/H_{P}$) &  1.5   &  1.6 / 1.0$^\dag$  &   2.3  &  1.68  & 2.0  &  1.63\\ 
Overshoot parameter  ($d/H_{P}$)    &   0.335 & 0.1 &0.2 &  $\sim 0.5$  & $f$= 0.0/{\bf 0.01}/0.02$^\ddag$ &  0.0/{\bf 0.1}/0.2\\
Rotation   &  0 - 550 \kms   & $\Omega/\Omega_{crit}=0.4$  &  300 \kms  &  0  & 0 / 200 \kms&   0 / 220  \kms \\
Magnetic field   &  Spruit-Taylor  & no  &  no   &  no   & no   & no \\
Solar mixture    &  AGS05$^6$ &  AGS05 & AGSS09$^7$  & GN93$^8$   & GN93  &  AGSS09 \\
   &   with C,N,O, &  with Ne enhanced  &  &  &&   \\
   &   Mg,Si,Fe modified & (Cunha et al. 2006)   &  &  &&   \\
\hline
\end{tabular}
\tablefoot{\newline References: 1 - \citet{brott11a}; 2 - \citet{ek12}; 3 - \citet{cl13}; 4 - \citet{bert09}; 5 - this work\\
Heavy elements solar mixture : 6 - \citet{asplund05} ; 7 - \citet{asplund09} ; 8 - \citet{gn93}\\
$\dag$ - For stars with initial mass $<$ 40 \msun, the mixing length parameter is $l/H_{P} = 1.6$. For more massive stars, it is defined  with respect to the local density scale height and $l/H_{\rho} = 1.0$.\\
$\ddag$ - In MESA the overshooting is implemented as a decreasing exponential with parameter $f$ (see text). }
\end{center}
\end{table*}

\subsection{STERN stellar evolution code \citep{brott11a}}
\label{s_stern}

We first make use of the grid of stellar evolution models published by
\cite{brott11a}. The computations have been performed with the code
fully described in \citet{hlw00}. In the following, we will refer to
this code as the STERN code.\\

\noindent {\bf Solar reference chemical composition}\\ 
\citet{brott11a} adopt tailored reference chemical abundances for
their models of LMC, SMC and Galactic massive stars based on the solar
abundances of \cite{asplund05} with a modification of the C, N, O, Mg,
Si and Fe abundances. This results in unusual chemical mixtures
described in Tables~1 and 2 of their paper. Their adopted values for
the metal mass fraction Z is 0.0088, 0.0047 and 0.0021 for the
Galaxy, the LMC and the SMC respectively. The Galactic metallicity is
about half the value used by the other codes (see below). The initial helium content is Y=0.264. The OPAL
radiative opacities of \citet{igro96} are used in the calculations. \\

\noindent {\bf Convection} \\
They use the Ledoux criterion\footnote{For a plasma described by a general equation of state, the {\em Ledoux stability criterion} is given by $\nabla_{rad} < \nabla_{ad} + \frac{\phi}{\delta} \nabla_\mu$ with $\delta = -\frac{ln \rho}{ln T}$ and $\phi = \frac{ln \rho}{ln \mu}$, $\mu$ being the mean molecular weight. When there are no chemical gradients, $\nabla_\mu = 0$ and the Schwarzschild stability criterion is recovered: $\nabla_{rad} < \nabla_{ad}$.} to determine the extension of the
convective regions, and model convection according to the mixing
length theory with $\alpha_{\rm MLT} = 1.5$. The
mixing length is the length over which a displaced element
conserves its properties, and the mixing length parameter $\alpha_{\rm MLT}$, is the
ratio of the mixing length to the local pressure scale height $H_P$. The zones which are stable according to the
Ledoux criterion but unstable according to the Schwarzschild criterion
are considered to be semi-convective. Semi-convection is included as in
\cite{langer83} with $\alpha_{\rm sc} = 1$. Finally Brott et al.\ calibrate an additional
classical overshooting parameter to adjust the evolution of the
rotation velocity as a function of surface gravity of a
16~\msun\ model at LMC metallicity (see Sect.\ \ref{s_ms}). This
parameter is applied to their entire grid and results in an extension
of the convective cores beyond the limit defined by the Ledoux
criterion by $d_{\rm over} = 0.335 H_p$.\\

\noindent {\bf Mass loss}\\
Mass loss is implemented following a combination of prescriptions or
recipes that are specific for each evolutionary phase of massive
stars. \cite{brott11a} use \cite{vink00,vink01} ($\dot{M}_{V}$) for winds of early O
and B-type stars. A switch to \mdot\ by \cite{ndj90} ($\dot{M}_{NdJ}$) is operated at \teff
$<$ 22000 K (bi-stability jump temperature) whenever \mdot$_{V}$ $<$ \mdot$_{NdJ}$. Another switch is
operated for the Wolf-Rayet phase, and \mdot\ by \citet{hamann95} is
adopted as soon as $Y_s \ge 0.7$ ($Y_s$ is the surface helium abundance). For intermediate values of $Y_s$, an
interpolation between the Vink et al. mass loss rates and the
Wolf-Rayet mass loss rates reduced by a factor 10 is performed. \\
Brott et al.\ use a metallicity scaling of the mass loss
by a factor $(Fe_{surf} / Fe_\odot)^{0.85}$ based on the solar iron
abundance from \cite{GNS96} ($\epsilon(Fe)=7.50$), which is higher than that of their
models ($\epsilon(Fe)=7.40$). For the rotating models, they also apply the correction
factor $\left(\frac{1}{1-V/V_{crit}}\right)^{0.43}$ to the mass loss rate
($V_{crit}$ is the critical velocity).\\

\noindent {\bf Rotation and rotation-induced mixing}\\
The effects of the centrifugal acceleration on the stellar structure
equations is considered according to \cite{kippen70}. The transport of
angular momentum and chemical species is treated in a diffusive way
following the formalism by \cite{es78} as described in
\cite{hlw00}. Eddington-Sweet circulation, dynamical and secular
shear, and axisymmetric (GSF) instabilities contribute to the
transport of both angular momentum and chemical species. The formalism they
use relies on two efficiency factors (free parameters) : $f_c =
0.0228$ which reduces the contribution to the rotation-induced
hydrodynamical instabilities in the total diffusion coefficient,
and $f_\mu = 0.1$ which regulates the inhibiting effect of chemical
gradients on the rotational mixing. The values of these parameters are
calibrated on observations. They are fully described in Eq.\ 53 and 54
of \citet{hlw00}.\\ 
In addition to that, the action of magnetic fields
on the transport of angular momentum \textit{only} is included
through the highly debated Tayler-Spruit dynamo \citep{spruit02}
following \citet{ply05}.

\subsection{Geneva stellar evolution code \citep{ek12}}

We also use the grid published by \cite{ek12}, and summarize briefly
the main physical ingredients used to compute the models provided in
that paper.\\

\noindent {\bf Solar reference chemical composition}\\ 
It is based on \citet{asplund05} with a modification of the Ne
abundance according to \citet{chl06}. Their adopted metals mass
fraction is $Z=0.014$ resulting from a solar calibration. The initial helium content is Y=0.266. The OPAL radiative opacities of \citet{igro96} are used. \\

\noindent {\bf Convection}\\ 
They use the Schwarzschild criterion to define the convective
regions. Convection is modelled following the mixing-length formalism
with $\alpha_{\rm MLT} = 1.6 $. For models with M $>$ 40 \msun\, the mixing length parameter is computed using the density height scale instead of the pressure height scale  with $\alpha_{\rm MLT} = 1$ following \citet{maeder87}.\\ 
They also include classical
overshoot at the convective core edge with $\alpha_{\rm over} =
0.1$. This parameter corresponds to the ratio between the extension of
the convective core beyond the value resulting from the Schwarzschild
criterion to the local pressure scale height.  $d_{\rm over} = 0.1 H_P$ is
calibrated to reproduce the width of the main sequence in the mass
range 1.35 - 9 \msun. Semi-convection is not modelled.\\

\noindent {\bf Mass loss}\\
Mass loss is implemented following a combination of
prescriptions or recipes chosen to best represent mass loss of
massive stars along their evolution. Ekstr{\"o}m et al.\ use the
stellar winds prescriptions from \citet{vink00,vink01} when log(\teff)
$> 3.9$. A switch to \cite{jager88} mass loss formulae is operated when
the models reach
log(\teff) $<$ 3.9 and then to \citet{paul00} as they evolve into the
red supergiant phase.  For the Wolf-Rayet phases (log(\teff) $>$ 4 and
$X_S$ $\leq$ 0.4 - $X_S$ being the surface hydrogen abundance), they use \citet{nl00} or
\citet{gh08}.\\ Mass loss rates are scaled
according to \citet{mm00b} for the rotating models.\\ 
For the most massive models (M $>$ 15\msun), in order to account for
the supra-Eddington mass loss during the red supergiant phase, they
multiply \mdot\ by a factor of 3 whenever the luminosity in the
envelope becomes 5 times larger than the Eddington luminosity. \\

\noindent {\bf Rotation and rotation-induced mixing}\\
The modification of the stellar structure equations by the centrifugal
acceleration is taken into account following \citet{mm97}. The
transport of angular momentum and of nuclides due to meridional
circulation and turbulent shear is self-consistently included
following the formalism by \citet{zahn92,mz98}. The prescriptions used
for the turbulent diffusion coefficients are from \citet{zahn92} for the
horizontal component and from \citet{maeder97} for the vertical
component. \\
Convective regions are assumed to rotate as solid-bodies.\\
No additional transport due to the presence of magnetic fields is
included.

\subsection{FRANEC stellar evolution code \citep{cl13}}
\label{s_franec}

\citet{cl13} reported on the inclusion of rotation in the FRANEC code and have computed a grid of massive stellar evolution models with and without rotation.\\

\noindent{\bf Solar reference chemical composition}\\
The models are computed using the heavy element solar mixture from
\citet{asplund09} . The initial global metallicity and helium content
are $Z=0.01345$ and $Y=0.265$.  The associated opacities are from OPAL
for radiative opacities.\\

\noindent{\bf Convection}\\ 
Convection is treated following the MLT formalism and convective
limits are defined using the Schwarzschild criterion except for the H
burning shell appearing at the beginning of the core He burning phase,
for which the Ledoux criterion is applied \citep[see][]{lcb03}. The
mixing length parameter adopted is not given, but should be of $\alpha
= \Lambda / H_p = 2.3$ according to \citet{scl97}. Classical
overshooting is included with a value of $d_{\rm over}$ = 0.2$H_p$.\\

\noindent{\bf Mass loss}\\
They use the mass loss prescriptions of \citet{vink00,vink01} for
the blue supergiant phase, switching to \cite{jager88} when log(\teff) $<$
3.9, and \citet{nl00} for the WR phase. The mass loss during the red
supergiant phase is enhanced according to \citet{vl05}. Mass loss of
rotating models is also enhanced as in the STERN code, following \citet{hlw00}.\\

\noindent{\bf Rotation and rotation-induced mixing}\\
The modification of the stellar structure equations by the centrifugal
acceleration and the
transport of angular momentum and of nuclides are the same as in the Geneva code.\\ 
The impact of the mean molecular weight
gradients on the transport of both angular momentum and nuclides is
regulated by the use of a free parameter $f_\mu$ defined by
$\nabla^{\rm adopted}_\mu = f_\mu \times \nabla_\mu$. Chieffi \& Limongi adopted
$f_\mu$=0.03. This value is calibrated to ensure that at solar
metallicity, the stars in the mass range 15-20~\msun\ settling on the
main sequence with an equatorial velocity of 300 \kms\ will increase
their surface nitrogen abundance by a factor of $\approx$ 3 by the
time they reach the TAMS.\\ 
Convective regions are assumed to rotate
as solid-bodies.\\ 
No additional transport due to the presence of
magnetic fields is included.\\

\subsection{Padova stellar evolutionary code \citep{bert09}}

The evolutionary models for massive stars computed with the Padova
code are described in \citet{bert09}. Additional information regarding
the code can be found in \citet{bono00} and
\citet{piet04,piet06}. Rotation is not included in the grid of
\citet{bert09}.\\

\noindent {\bf Solar reference chemical composition}\\
The metals mass fraction adopted for their solar-metallicity models is
$Z=0.017$. We used the models with Y=0.26 (several values are
available). The OPAL radiative opacities of \citet{igro96} are used. A
global scaling of the relative element mass fractions is made compared
to the mixture of \citet{gn93} on which the OPAL tables are based. For
very high temperatures ($\log T >$ 8.7), the opacities of
\citet{weiss90} are used.\\

\noindent {\bf Convection}\\
They adopt the formalism of the mixing length theory with
$\alpha_{MLT} = 1.68$, calibrated on solar models. Stability is set by
the Schwarzschild criterion. Overshooting is taken into account with a free parameter corresponding to
the ratio of the true extent of the convection core to the convection
core radius defined by the Schwarzschild criterion. The notable
difference in the Padova code is that this parameter expresses the
extent of overshooting \textit{across} (and not above) the border of the
convective core (as set by the Schwarzschild criterion). A value of
0.5 for the overshooting parameter is adopted. Overshooting below the
convective envelopes is also accounted for with a parameter equal to 0.7.\\
Semi-convection is considered to be negligible for massive stars models.\\

\noindent {\bf Mass loss}\\ 
The mass loss prescriptions of \citet{jager88} are used for all phases
of evolution. A metallicity scaling of radiatively driven winds is
taken into account according to \citet{kud89} (i.e. $\dot{M} \propto
Z^{0.5}$).

\subsection{STAREVOL \citep{dmp09}}

A detailed description of the STAREVOL code can be found in
\citet{sdf00,siess06,dmp09}. The models computed for the present study
were obtained using the STAREVOL v3.30 which includes a number of
updates with respect to previous descriptions of the code\footnote{The
  updates concern opacities and reference solar abundances, as well as
  mass-loss prescriptions.}. For the present study we have adopted a
setup close to that used in the Geneva grid.\\

\noindent {\bf Solar reference chemical composition}\\ 
We use \citet{asplund09} with OPAL tabulated opacities modified
accordingly. At low temperature we use the \citet{fergu05} opacities
computed for the Asplund et al. 2009 solar composition. The adopted
solar metallicity is thus $Z=0.0134$. The initial helium content is
$Y=0.277$. No further modification of the abundances is made. When
computing models for non-solar metallicities, a simple proportionality
is applied.\\

\noindent {\bf Convection}\\ 
The Schwarzschild criterion is used to define the convective
regions. Convection is modelled following the mixing-length formalism as
described in \citet{kw90}. The mixing length parameter $\alpha_{\rm
  MLT} = 1.64$ is calibrated for the solar model. In some models, we have included classical overshoot at the edge of convective regions with $\alpha_{\rm over}$ = 0.1 or
0.2.\\

\noindent {\bf Mass loss}\\
 For massive stars (M $>$ 7~\msun)
with log(\teff) $> 3.9$ we apply the prescriptions
from \citet{vink00,vink01}, which we change to (a) \cite{jager88} when
the models evolve to the red and have their temperature drop below
log(\teff) = 3.9 and then to \citet{paul00} as they evolve into the
red supergiant phase, (b) to \citet{r75} for models in the mass range
7-12~\msun\ when they evolve off the main sequence, (c) to
\citet{nl00} for those models that experience a Wolf-Rayet phase
(e.g. with log(\teff) $>$ 4 and $X_{S}$ $\leq$ 0.4).\\ 
The mass loss is down-scaled by a factor $(Z/Z_\odot)^{0.5}$ for
non-solar metallicity models. We have included the correction to the mass loss of rotating
massive stars according to \citet{mm01}.\\

\noindent {\bf Rotation and rotation-induced mixing}\\
The modification of the stellar structure equations due to centrifugal
acceleration in rotating models is taken into account following
\citet{kippen70}. The expression for the effective temperature
following this formalism is implemented as described in Appendix A of
\citet{mm97}. In addition to this, the transport of angular momentum
and of nuclides is as in the Geneva and FRANEC codes. The
prescriptions used for the turbulent diffusion coefficients are from
\citet{zahn92} for the horizontal component and from \citet{tz97} for
the vertical component.\\ 
Convective regions are assumed to rotate as
solid-bodies.\\ No additional transport due to the presence of
magnetic fields is included.\\

\subsection{MESA code \citep{pax11}}

MESA\footnote{http://mesa.sourceforge.net/} is a public distribution
of modules for experiments in stellar astrophysics. The computation of
evolutionary models is possible with the module ``star'' of the
distribution. An exhaustive description of the code is available in
\citet{pax11} and \citet{pax13}. We have computed dedicated
evolutionary models for 7, 9, 15, 20, 25, 40 and 60 \msun\ stars. Both
non-rotating and rotating (initial equatorial velocity of 200 \kms)
models have been calculated.\\

\noindent {\bf Solar reference chemical composition}\\
We have adopted a value of $Y=0.26$ and $Z=0.014$ in our
calculations. MESA uses OPAL opacities from \citet{igro96}. The
relative mass fraction of metals in the OPAL composition is based on
the solar composition of \citet{gn93}, \citet{gs98} or
\citet{asplund09}. The user can select any of these compositions. A
global scaling with Z is made when non-solar metallicity models are
computed.\\

\noindent {\bf Convection}\\
The standard mixing length formalism as defined by \citet{cox68} is
used to treat convection as a diffusive process in MESA. The onset of
convection is ruled by the Schwarzschild criterion. In our
calculations, we used $\alpha_{MLT} = 2.0$. Although it is available,
we did not include semi-convection in our models. Convective
overshooting is treated as a diffusive process following the formalism
of \citet{herwig00}. The overshooting diffusion coefficient ($D_{ov}$)
is related to the MLT diffusion coefficient ($D_{conv}$) through
$D_{ov}=D_{conv} e^{-\frac{2z}{fH_{p}}}$ where $f$ is a free
parameter. Unless stated otherwise, we have adopted $f=0.01$ in our
calculations.\\

\noindent {\bf Mass loss}\\
A mixture of prescriptions is used to account for mass loss in the
various phases of evolution. The recipe of \citet{vink01} is
used for $\teff\ > 10000$K and $X(H)>0.4$. For the same temperature
range, but lower H content ($X(H)<0.4$), the mass loss rates of
\citet{nl00} are implemented. For $\teff\ < 10000$K, the values of
\citet{jager88} are used. It is possible to scale these prescriptions
by a constant factor. For hot star, it is a way to take the
metallicity dependence of mass loss rates into account (see
Sect.\ \ref{s_effect_ingredients}).\\

\noindent {\bf Rotation and rotation-induced mixing}\\
The geometrical effects of rotation are implemented following the
formalism of \citet{kippen70}.  The transport of angular momentum and
chemical species through meridional circulation and hydrodynamical
instabilities turbulence is treated as a purely diffusive process,
following the \citet{es78} formalism as in the STERN code. The
efficiency factor (see Sect.\ \ref{s_stern}) have the following
values: $f_c = 1/30$, similar to the theoretical value of
\citet{cz92}, and $f_\mu = 0.1$.  We did not include magnetism in our
computation (although the formalism of \cite{spruit02} is implemented
in MESA and can be switched on).


\section{Code predictions and uncertainties}
\label{s_comp_theo}

In this section, we perform comparisons between the results of
calculations performed with the six codes described above. We focus
on the evolutionary tracks in the Hertzsprung-Russell diagram. We first
compare standard tracks (i.e. without rotation) in order to test the
various implementations of the basic physics. We subsequently
investigate the effects of rotation.

\subsection{Effects of physical ingredients on evolutionary tracks}
\label{s_effect_ingredients}

\begin{figure}[t]
     \centering
          \includegraphics[width=6.5cm,height=6cm]{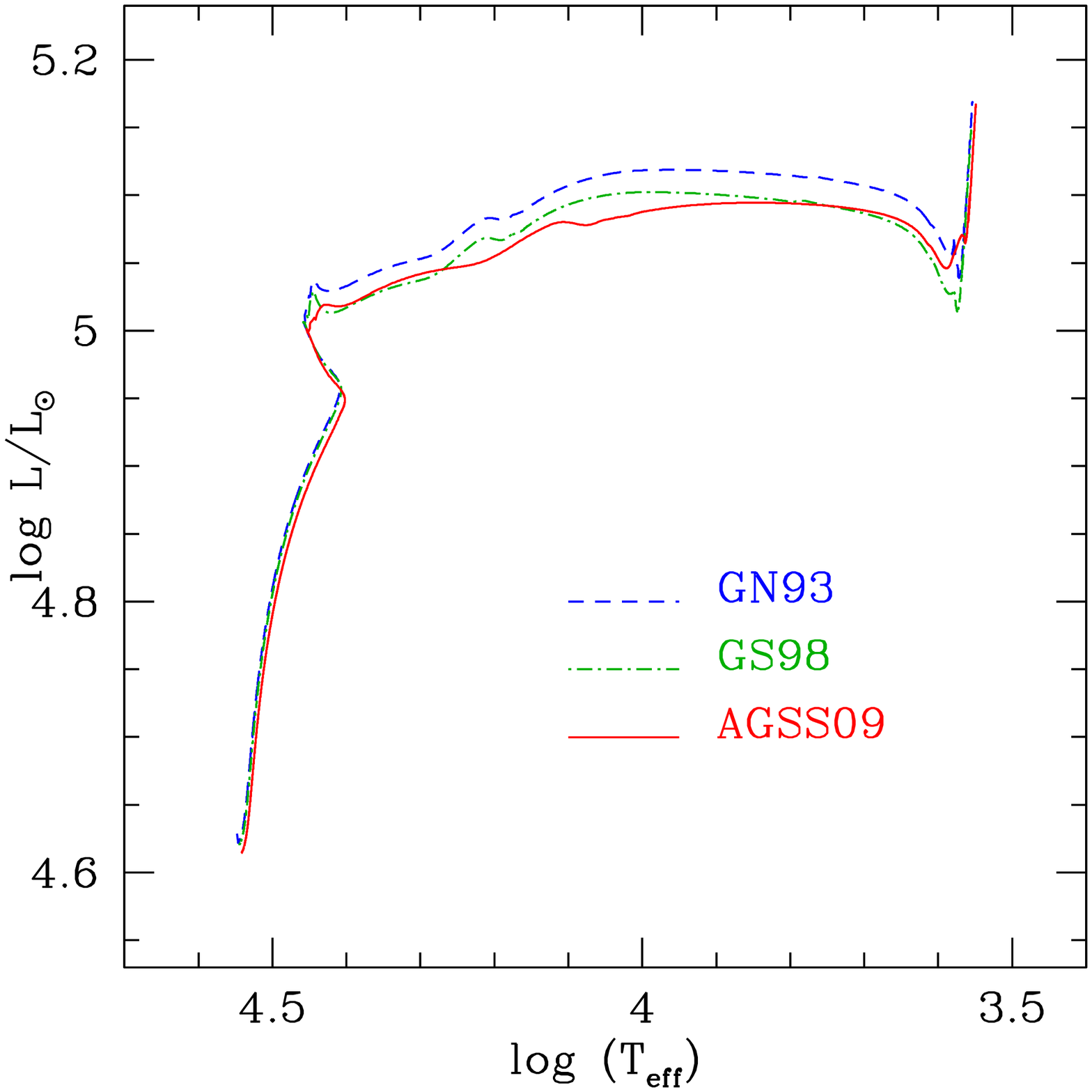}
          \includegraphics[width=6.5cm,height=6cm]{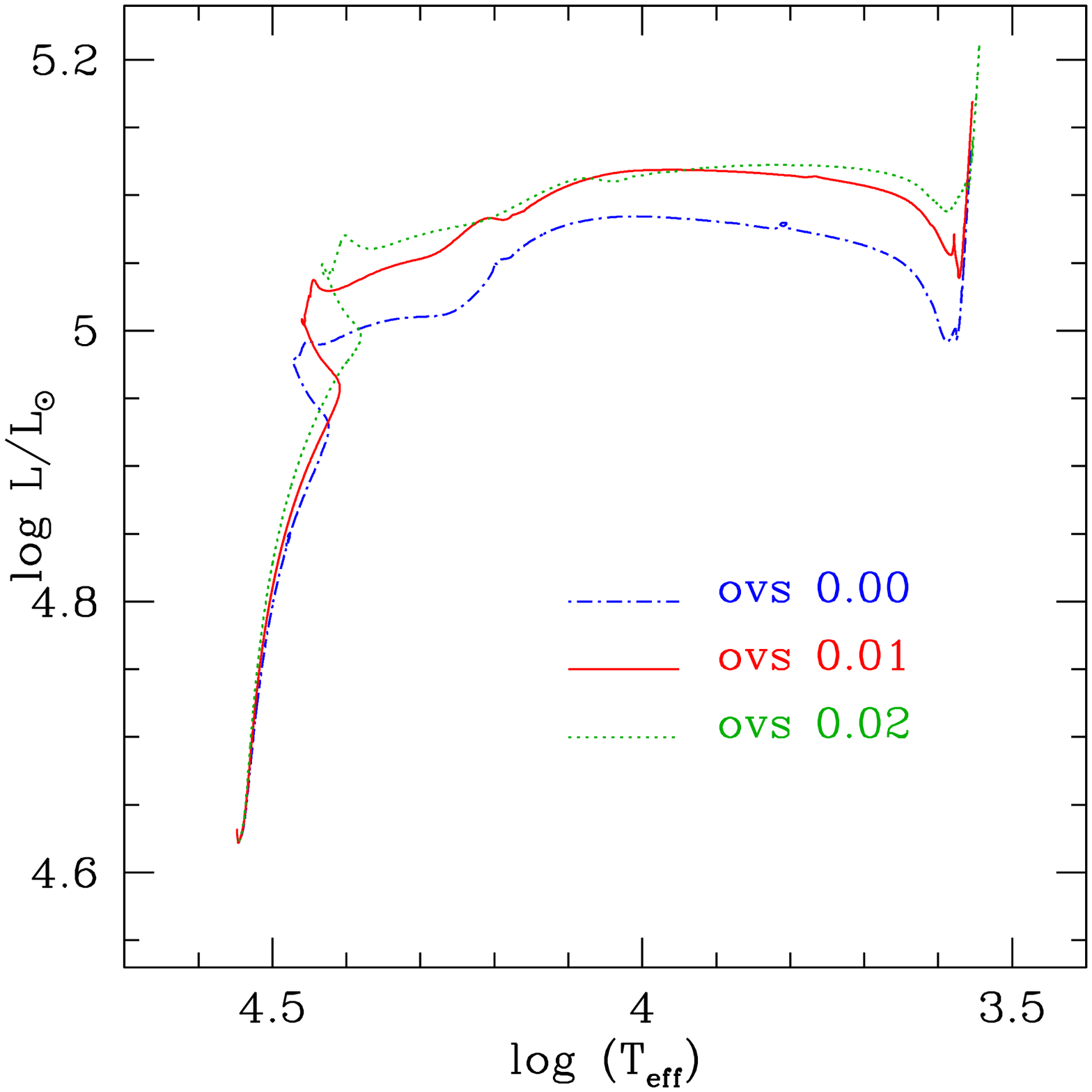}
          \includegraphics[width=6.5cm,height=6cm]{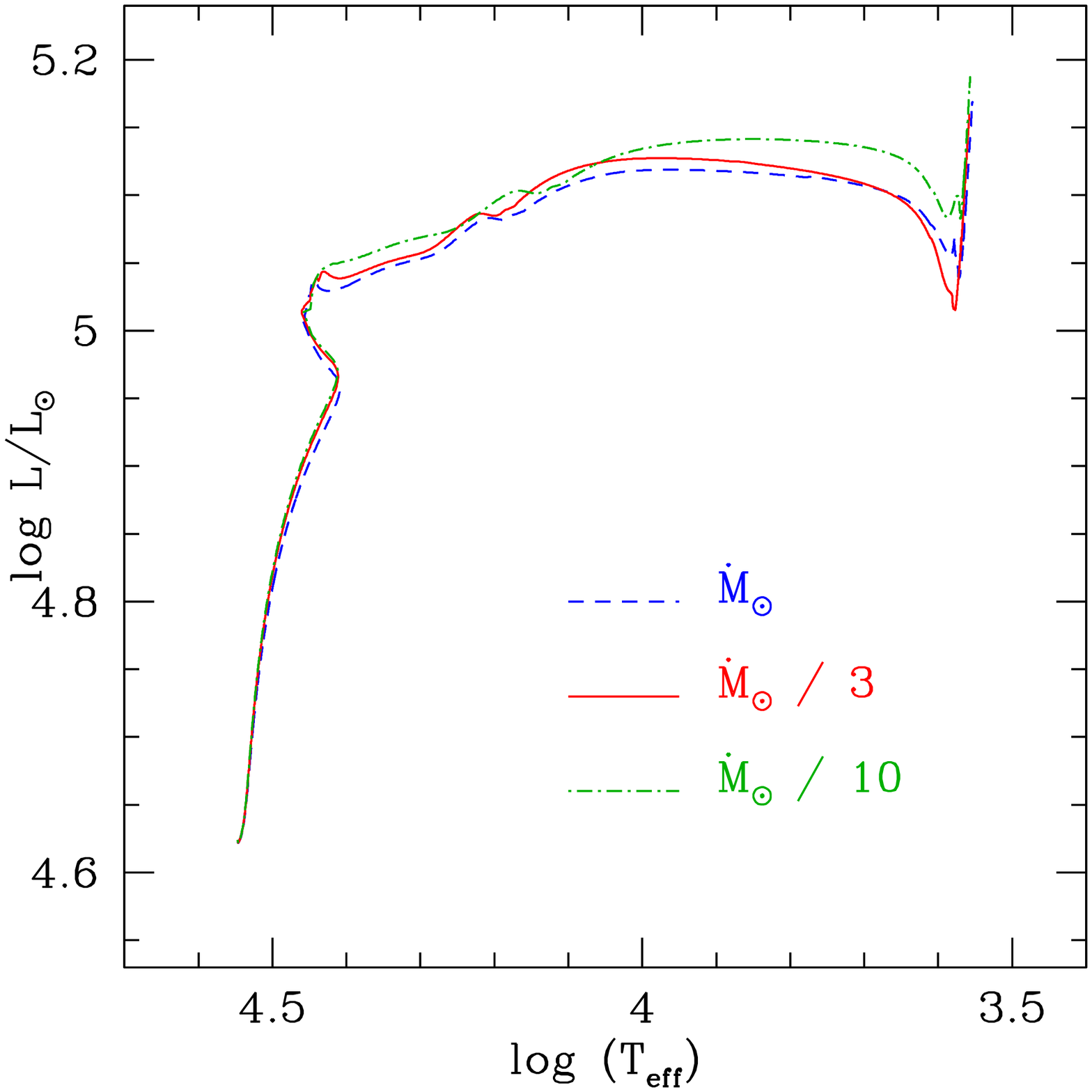}
     \caption{Effects of opacities (top), overshooting (middle) and mass loss (bottom) on 20 \msun\ evolutionary tracks. The computations have been performed with the code MESA. In the upper right panel, different solar heavy elements mixtures are used, for the same initial metal fraction ($Z=0.014$). GN93 refers to \citet{gn93}, GS98 to \citet{gs98} and AGS09 to \citet{asplund09}. }
     \label{fig_effects}
\end{figure}

In Fig.\ \ref{fig_effects} we illustrate the effect of modifying the opacities, the overshooting and the mass loss on the evolution of a $M=20$ \msun\ model from the Zero Age Main Sequence (ZAMS) to the Terminal Age Helium Main Sequence (TAHeMS). We have used both the MESA and STAREVOL codes to compute the evolutionary sequences. We mainly discuss the MESA models in this section (the results obtained using STAREVOL are very similar).\\
The top panel illustrates the effect of different heavy elements solar mixtures on the opacities. The solar composition of \citet{gn93} and \citet{asplund09} are relatively similar: the C, N, O and Ne abundances do not differ by more than $\sim$0.05 dex. The \citet{gs98} abundances are on average 0.10-0.15 dex larger. In Fig.\ \ref{fig_effects}, we see that the main effects of different solar composition on the opacities is reflected in the post main sequence evolution. On the main sequence, the luminosity variations are negligible, while beyond the terminal age main sequence (TAMS) the differences vary between 0.01 and 0.02 dex depending on the temperature. 

The middle panel of Fig.\ \ref{fig_effects} shows the effect of overshooting. The calculations have been performed for a diffusive overshooting with a parameter $f$ equal to 0, 0.01 and 0.02 as indicated on the figure. Overshooting is included only in the convective regions related to H burning. Qualitatively, the effect of an increasing overshooting is the lengthening of the main sequence phase. As a result of the larger extension of the convective core, a larger amount of hydrogen is available for helium production in the core. Quantitatively, the main sequence duration is 8.60 Myr for $f$=0.01 and 9.06 Myr for $f$=0.02. This corresponds to an increase of 9\%. As a consequence of the longer main sequence duration for larger overshooting, the star exits the core H burning phase at a lower effective temperature (by 2500 K), and at a higher luminosity (increase of 0.05 dex) when $f$ increases from 0.01 to 0.02. If the overshooting parameter is not constrained, a degeneracy in the evolutionary status of a star located close to the end of the main sequence can appear. Depending on the tracks used and the amount of overshooting, it can be identified as a core H burning object close to end of the main sequence, or as a post main sequence object. Beyond the main sequence, models with stronger overshooting evolve similarly but at higher luminosities.

The bottom panel of Fig.\ \ref{fig_effects} illustrates the effect of mass loss rates on evolutionary paths. In addition to the track with the standard mass loss rate, two additional tracks with mass loss rates globally scaled by a factor 0.33 and 0.10 are shown. As expected, the main sequence is barely affected. The reason is the low values of the mass loss rates during this phase for the initial mass of 20 \msun\ considered here. For the standard track (dashed blue line) the mass at the end of the main sequence is 19.66 \msun, corresponding to a loss of only 1.7\% of the initial mass over 8.60 Myr. The mass drops to 18.26 \msun\ in the next Myr (time to reach the bottom of the red giant branch). On average, the mass loss rate is thus 35 times larger in the post-main sequence phase compared to the main sequence. To first order, the effect of mass loss can be understood as a simple shift to lower luminosity. Since the luminosity is directly proportional to some power-law of the mass \citep[the exponent being around 1.0-2.0 depending on the mass, e.g.][]{kw90}, a reduction of the mass immediately translates into a reduced luminosity. This is what we observe in Fig.\ \ref{fig_effects}. Quantitatively, a reduction by a factor close to 3 (10) in the mass loss rates corresponds to a maximum increase in luminosity of $\sim$0.01 (0.03) dex. The changes are larger for more massive stars since mass loss rates are also higher.

The prescriptions of mass loss rates for massive stars suffer from several uncertainties. The presence of clumping in hot stars winds has lead to a reduction of the mass loss rates by a factor of roughly 3 \citep{puls08}. But this value is still debated, reduction up to a factor of 10 being sometimes necessary to reproduce observational diagnostics \citep{jc05,full06}. For the cool part of the evolution of a massive star, the very nature of the mass loss mechanism is still not clear. \citet{mj11} have shown that the mass loss rates of \citet{jager88} are still valid. But for a given luminosity, the scatter in mass loss rates is large (up to a factor 10). The uncertainties in the mass loss rates thus translate into uncertainties of the order of 0.02 dex in the luminosity of evolutionary tracks beyond the TAMS.

\begin{figure}[t]
\centering
\includegraphics[width=.45\textwidth]{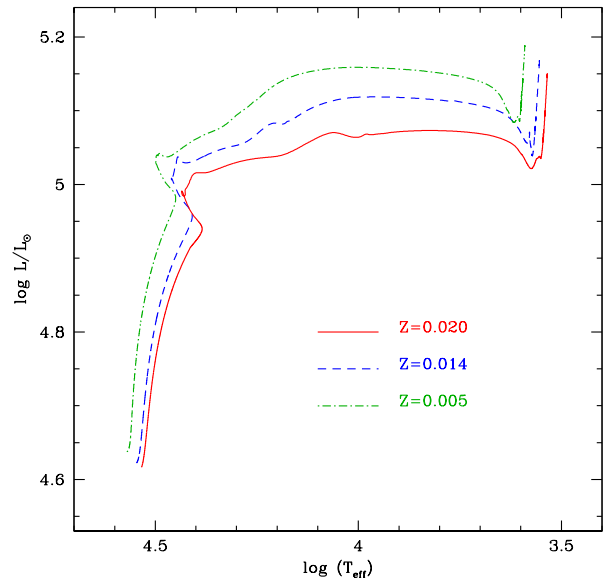}
\caption{Effect of metallicity on a 20 \msun model computed with MESA.}
\label{figZ}
\end{figure}

Figure\ \ref{figZ} highlights the well documented effects of metallicity \citep[e.g.][]{mm03}. We have computed models for three different metallicities: the solar value ($Z=0.014$) and the extreme values encountered in the Galaxy according to the study of HII regions by \citet{balser11} -- $Z = 1.5~Z_{\odot}$ and $Z = 1/2.5~Z_{\odot}$. No scaling of the mass loss rates was applied in order to extract the effect of metallicity on the internal structure and evolution. A lower metal content corresponds to a lower opacity, which in turn translates into a higher luminosity. On average, a reduction of the metal content by a factor of two translates into an increase in luminosity by 0.005-0.010 dex on the main sequence and by 0.03-0.05 dex beyond.

Assuming a typical uncertainty on the luminosity of $\pm$ 0.02 dex (opacity effect), $\pm$0.04 dex (overshooting effect), $\pm$0.01 dex (mass loss effect), $\pm$0.03 dex (metallicity effect) and simply adding quadratically the errors, we obtain a global uncertainty of about $\pm$0.05 dex on the luminosity of a MESA track. The values we adopted are typical of the uncertainties at the end of the main sequence and around \teff\ = 10000 K. An uncertainty of 0.05 dex on the luminosity is equivalent to an uncertainty of about 6\% on the distance of the star. On the main sequence, the uncertainty on the luminosity is lower than $\pm$0.02 dex. \\ 

\begin{figure*}[]
     \centering
     \subfigure{
          \includegraphics[width=.45\textwidth]{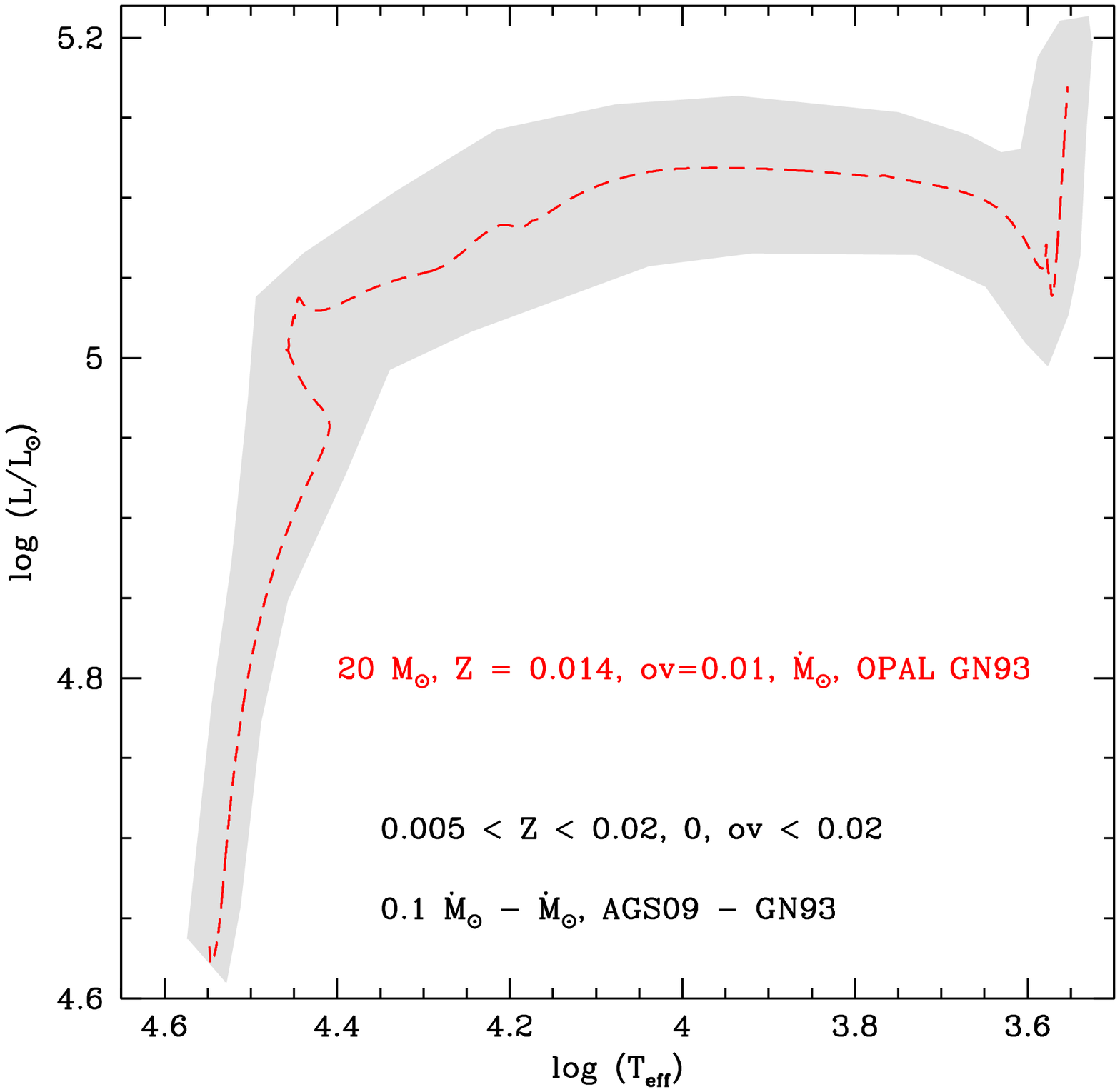}}
     \hspace{0.2cm}
     \subfigure{
          \includegraphics[width=.45\textwidth]{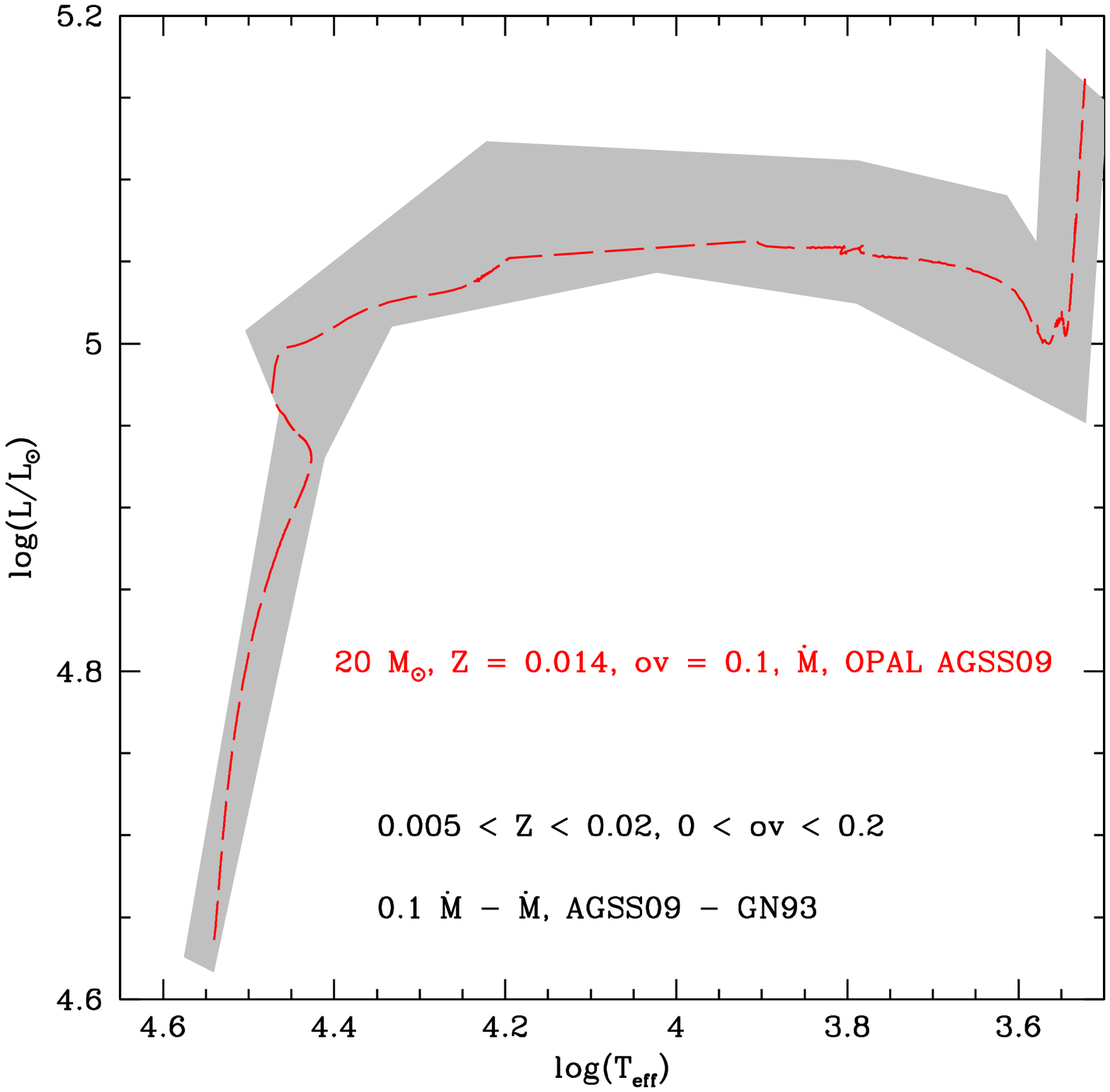}}
     \caption{Region occupied by the evolutionary tracks of 20 \msun\ models computed with MESA (left) and STAREVOL (right) with different opacities, metallicities, mass loss rates and overshooting parameters. The dashed red line shows the track of the standard model, with the parameters as indicated on the figure (the overshooting parameter - ov - is not defined in the same way in both codes, hence the different values). The shaded envelope defines a rough global intrinsic uncertainty on 20 \msun models.}
     \label{fig_shade}
\end{figure*}

We have computed a second set of these models using the STAREVOL code, and we also find that the choice of the overshooting and of the metallicity are the ones affecting the most the luminosity. The global uncertainty on the 20 \msun\ track computed with STAREVOL is of $\pm$ 0.06 dex around \teff\ = 10000 K, of the same order than that found for MESA models. In Fig.~\ref{fig_shade}, we display the envelope corresponding to the global intrinsic error for the MESA and STAREVOL models of 20 \msun. For both sets of models, the shape of the envelope is similar. The uncertainty is maximum at temperatures around 10000 K (in the core He burning phase, see Table \ref{tab_param}). The uncertainty on the luminosity at a given effective temperature is not symmetrical with respect to the reference track shown in dashed line. These envelopes illustrate the intrinsic uncertainties of a given evolutionary track. \\

We have performed the same type of calculations and comparisons on a $M=7$ \msun\ model. The effect of mass loss is negligible (changes in luminosity smaller then 0.01 dex). Different chemical mixtures and their effect on opacities translates to uncertainties $<$ 0.05 dex on the luminosity. They are roughly similar to the effects seen in the $M=20$ \msun\ star. A change in metallicity from solar to 1/2.5 solar corresponds to an increase in luminosity by 0.1 dex. Hotter temperatures are also obtained. The effect is larger than in the M=20 \msun\ model. Finally, the largest effect on the evolutionary track of the $M=7$ \msun\ model is due to changes in the overshooting parameter. Variations in luminosity by 0.10-0.15 dex are observed beyond the main sequence. The effect of overshooting dominates the uncertainty on the $M=7$ \msun\ evolutionary track. The global uncertainty on the luminosity of the $M=7$\msun\ model amounts to 0.2 dex, which equivalent to an error of 30\% on the distance.

\subsection{Comparison between codes}
\label{s_comp_codes}

\begin{figure*}[t]
     \centering
     \subfigure{
          \includegraphics[width=.45\textwidth]{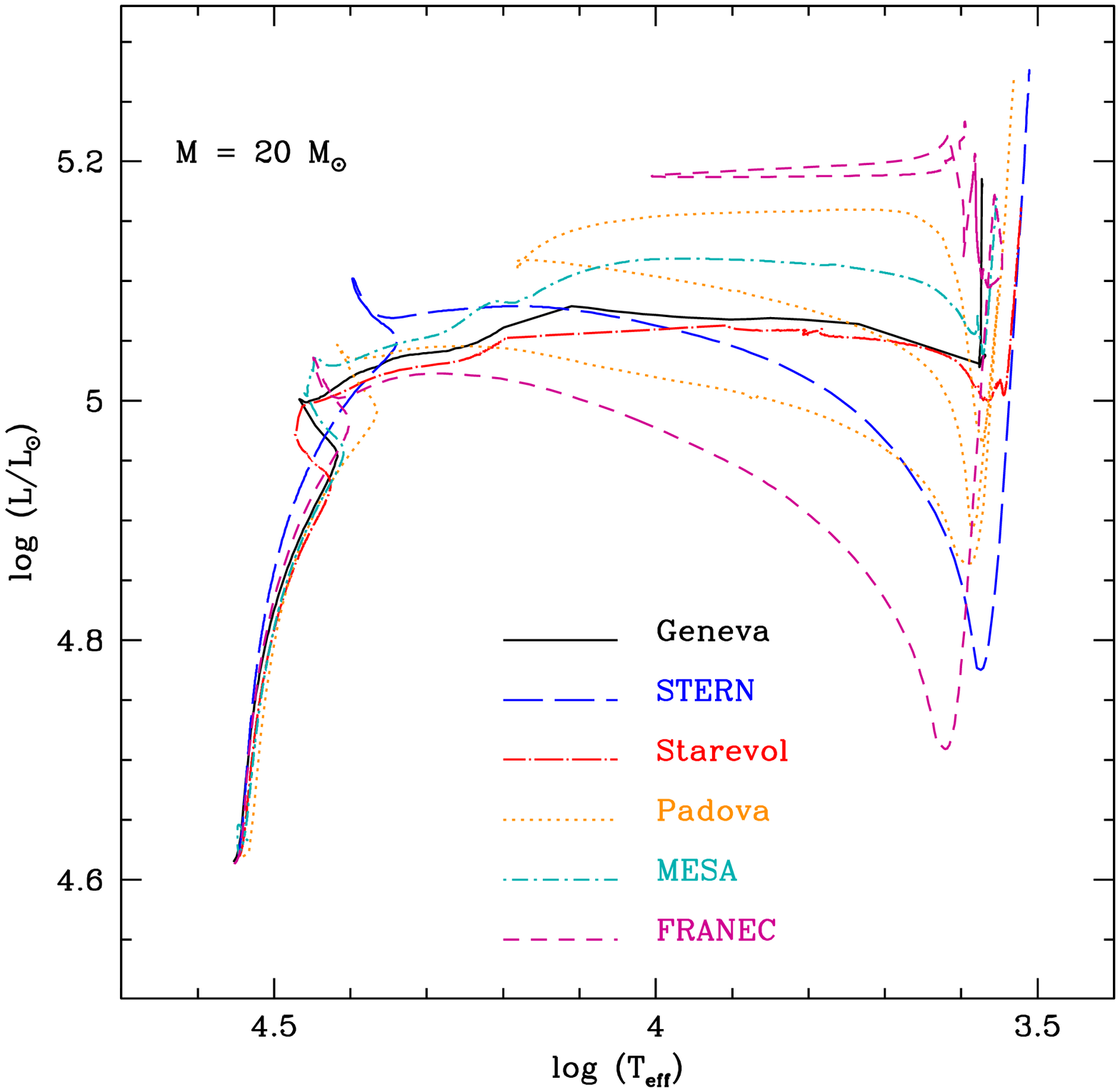}}
     \hspace{0.2cm}
     \subfigure{
          \includegraphics[width=.45\textwidth]{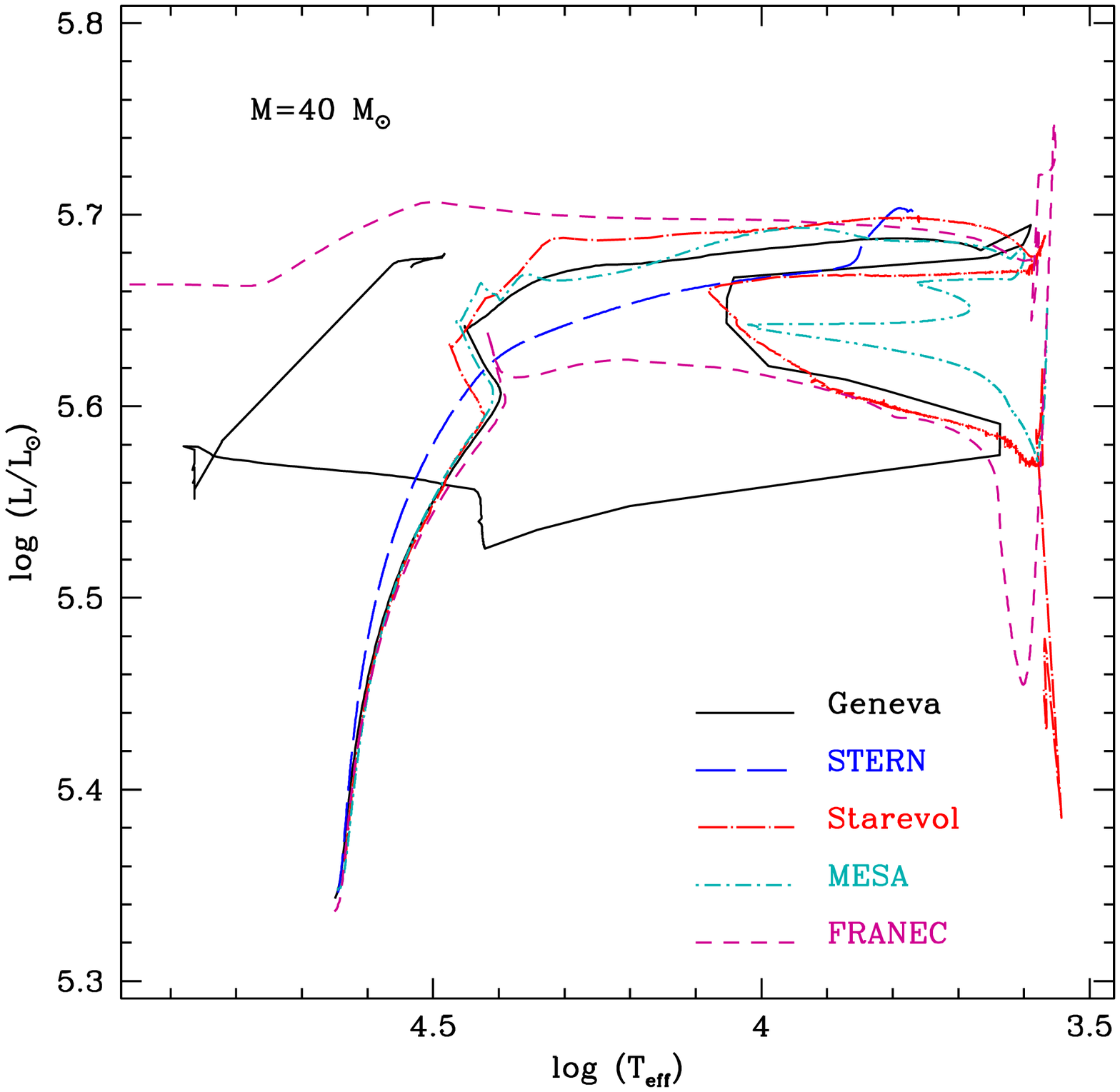}}
     \caption{Evolutionary tracks for $M=20$ \msun\ (left) and $M=40$ \msun\ (right), without rotation. For the $M=40$ \msun\ case, no Padova track exists.}
     \label{fig_Mcodes}
\end{figure*}

In this section we make direct comparisons between the six codes presented in Sect.\ \ref{s_models}. We focus on the $M=20$ \msun\ track corresponding to a late O star on the main sequence. Table \ref{tab_param} gathers the effective temperature, luminosity and age at four different evolutionary phases, for the six types of models. The 40 \msun\ track is also briefly presented at the end of this section.

Fig.\ \ref{fig_Mcodes} (left) shows the evolutionary tracks of classical (no rotation included) 20 \msun\ model at solar metallicity. On the main sequence there is an overall good agreement between the outputs of the different codes.
The STERN track is overluminous and bluer than the others which is expected from its lower metallicity ($Z = 0.0088$ vs $Z = 0.014 - 0.017$ for all the other tracks). 
The TAMS of the Geneva, STAREVOL and MESA models is located in the same region of the HR diagram as can also be verified from Table~\ref{tab_param}. However, the age at the TAMS is quite different for the Geneva Model, which is younger by $\approx$ 0.7 Myr compared to the STAREVOL and MESA models. A difference in age at the TAMS with no associated difference in \teff\ nor luminosity may indicate differences in the nuclear physics.\\ 
The characteristic hook at the end of the main sequence occurs at lower temperatures for the FRANEC, Padova and STERN models. The larger amount of overshooting, hence the size of the H core, is responsible for these differences (see Table \ref{tab_mod}). From Table~\ref{tab_param}, we also see that FRANEC and Padova models reach the TAMS later (by $\sim$0.5 Myr) compared to MESA and STAREVOL models, which is consistent with having a larger reservoir of fuel to be consumed during the hydrogen core burning phase. On the other hand, and surprisingly, the STERN model is even younger than all the models except the Geneva one, when it reaches the TAMS. The lower metal content might lead to a higher core temperature and consequently to a faster hydrogen burning via the CNO reactions. At the end of the main sequence, an age spread of 1.3 Myr (15\%) is observed between the six codes. 

\begin{figure}[]
\centering
\includegraphics[width=.45\textwidth]{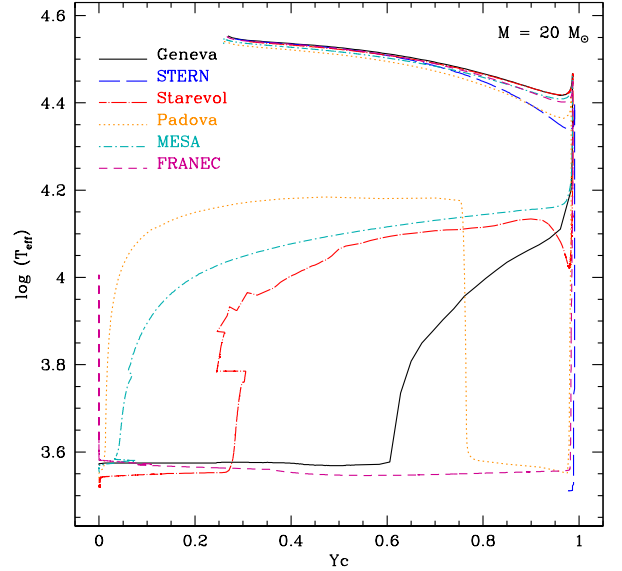}
\caption{Effective temperature as a function of central helium mass fraction for a $M=20$ \msun\ model computed with the six codes considered in this study.}
\label{teff_Y}
\end{figure}

Beyond the main sequence, the differences between the various codes are larger. The Geneva/Starevol/MESA tracks are still rather similar in the post main sequence evolution, with differences in luminosity usually lower than 0.05 dex. The red supergiant part of the STAREVOL model is located redwards compared to that of the MESA and Geneva models. The reason for this behaviour is attributed to a combination of differences in the mixing length parameter, the opacities and the equation of state.
The Padova and FRANEC models reach the TAMS with larger luminosities due to the stronger overshoot, and see their luminosity subsequently decrease to a large amount (0.3 dex in the case of the FRANEC model). This behaviour is also observed in the STERN model. The Padova, STERN and FRANEC models reach \lL = 4.87, 4.77 and 4.71 respectively at the bottom of the red giant branch, compared to the \lL = 5.00--5.05 reached by the MESA, STAREVOL and Geneva models. The decrease of the total luminosity during the Hertzsprung gap results from a subtle balance between the core contraction, the energy generation by the H shell surrounding the core, the mean molecular weight gradient profile and the opacity of the surface layers. For the Padova, STERN and FRANEC models, the thermal instability of the envelope (triggered by the above conditions) seems to be stronger, leading to a larger overall reduction of the luminosity. Given that the detailed structures associated with these tracks are not all available, it is difficult to be more precise concerning the different paths followed by the tracks presented here.\\
The beginning of the helium core burning phase (ZAHeMS) is defined as the time at which the central helium mass fraction starts to decrease from its maximum value reached after the central hydrogen burning phase. The ZAHeMS starts at \teff\ $\sim$ 25000 K for most models, except the Padova (\teff\ = 15657 K) and the STERN (\teff\ = 5388 K) ones. We attribute the very different temperature of the STERN models to the large overshooting and the inclusion of magnetism. The temperatures at the TAHeMS differ by 350 K at most. This is a large difference (10\%), affecting the interpretation of the properties of red supergiants \citep[e.g.][]{levesque05,davies13}.
The tracks from FRANEC and Padova tracks present a blue hook similar to what is observed during core helium burning for lower masses. \\
The large differences in the post main sequence evolution can
  also be seen in Fig.\  \ref{teff_Y} where we show the
  evolution of effective temperature as a function of core helium mass
  fraction (Y$_C$). The MESA and STAREVOL tracks spend most of the
  helium burning phase at temperatures larger than 10000 K. 40 \% of
  the helium burning phase takes place at hot temperatures in the Geneva model. The subsequent evolution takes place mainly at low \teff. In the FRANEC model, almost all the helium burning is done in the cool part of the HR diagram. Finally, the Padova model features a blue loop so that helium burning is first done at low \teff\ before finishing at \teff\ $>$ 10000 K. 

To summarize, the evolution of the $M=20$ \msun\ model becomes more uncertain as temperature decreases (i.e. as the star evolves), with a wider spread in luminosity in the HR diagram. This type of differences also exists for lower mass stars not analyzed here\footnote{See for instance the results presented at the workshop " The Giant Branches" held in Leiden in May 2009 - http:///www.lorentzcenter.nl/lc/web/2009/324/program.php3?wsid=324
.}, and reveals the uncertainty of the He-burning phases understanding and modelling.  The length of the main sequence depends on the treatment of overshooting, as illustrated in Sect.~\ref{s_effect_ingredients}. Differences in luminosity up to 0.3 dex (a factor of 2) and in temperatures up to 10\% are observed in the coolest phases of evolution.\\

\begin{figure}[t]
\centering
\includegraphics[width=9cm]{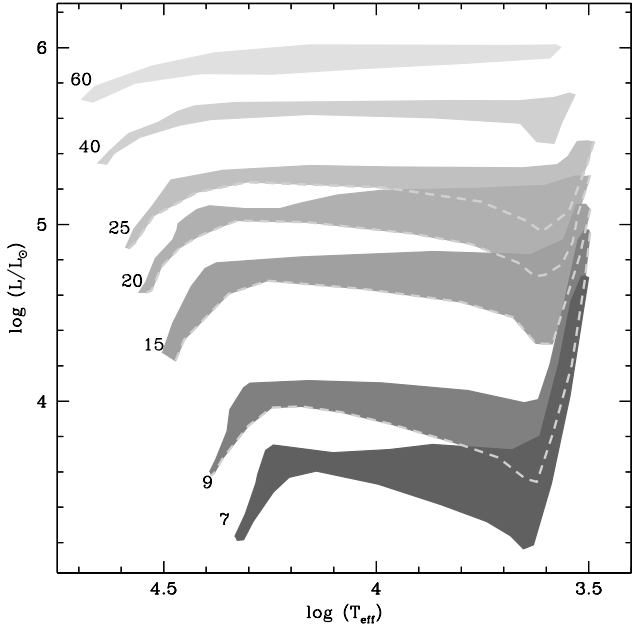}
\caption{Envelopes of evolutionary paths for $M=$7, 9, 15, 20, 25, 40 and 60 \msun\ taking into account the predictions of the codes studied in this paper. Rotation is not included. The envelopes for the 7 and 9 \msun\ models do not include FRANEC since no tracks is available for these masses. The Padova models do not exist for 40 and 60 \msun. For the 40 and 60 \msun\ tracks, the envelope encompasses only the redward evolution (the Wolf-Rayet phases are not included).}\label{fig_env_all}
\end{figure}

\begin{figure}[t]
\centering
\includegraphics[width=9cm]{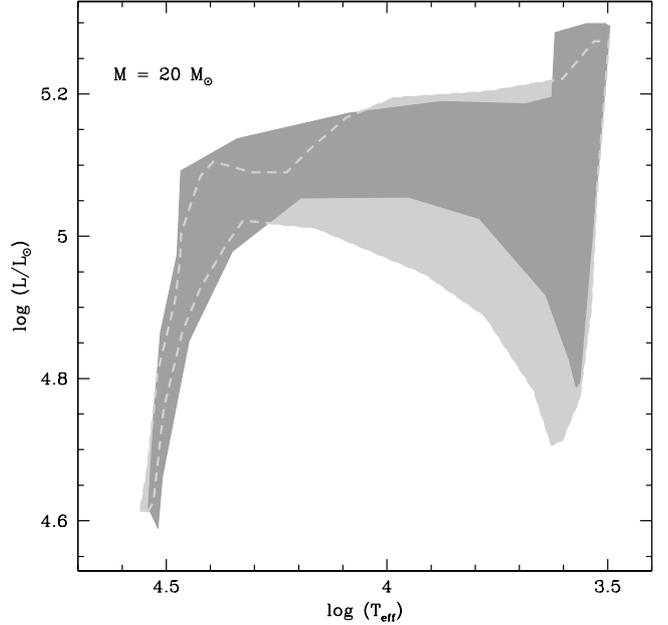}
\caption{Uncertainty on the location of the evolutionary path for a 20 \msun\ stellar model with (dark grey envelope) and without (light grey delimited by dashes lines) rotation. We have considered tracks generated by five different codes (Geneva, STERN, FRANEC, MESA and Starevol) with similar (yet not exactly the same) initial rotation rates.}\label{fig_env_rot}
\end{figure}

Fig.\ \ref{fig_Mcodes} (right) shows the evolutionary tracks for a
$M=40$ \msun\ star. Padova models for such a mass are not available
(the grid of Bertelli et al.\ 2009 stops at $M=20$ \msun). The Geneva,
MESA and STAREVOL models are very similar during the H and He burning
phases. The FRANEC track has the same path as the Geneva/MESA/STAREVOL
tracks on the main sequence. Beyond that, its luminosity is about 0.05
to 0.08 dex lower until it reaches the red part of the HR diagram
where it drops by another 0.12 dex before rising again by 0.25 dex and
starting its blueward evolution towards the Wolf-Rayet phase. The
behaviour is very different from the other models since once in the
red part of the HR diagram, the luminosity {\em increases} instead of
{\em decreasing}. This behaviour is tentatively attributed to
  the size of the H-rich envelope that develops in the FRANEC model
  when the luminosity drops. This envelope reaches 15 \msun\ at the
  lowest effective temperature. For comparison, in the MESA model it
  represents at most about 10 \msun. The STERN track is more luminous than all other tracks on the main sequence. It does not show the characteristic hook revealing the end of the convective core H burning phase. This feature is very peculiar in the STERN track and is attributed to the combination of the very large overshooting parameter and the inclusion of magnetism (see Sect.\ \ref{s_ms}). \\

In Fig.\ \ref{fig_env_all} we present a summary of the comparison between the available standard tracks. For each mass, we present envelopes encompassing the tracks produced by the six codes. These envelopes are defined from the ZAMS to temperatures of about 3000 K. For sake of clarity, the subsequent evolution (back to the blue) of the most massive objects (M$>$40 \msun) is not taken into account to create these envelopes. Such envelopes provide a first guess of the uncertainty on the evolutionary tracks for specific masses. The main sequence phase is relatively well defined with the major uncertainties being encountered at the exhaustion of central hydrogen, when the track makes a hook in the HR diagram. For the masses shown in Fig.\ \ref{fig_env_all}, there is no overlap. The same conclusion remains for the post main sequence evolution above 40 \msun. For the lower masses there is a degeneracy on the mass at low temperatures where the envelopes defining the possible location for a specific mass overlap. The overlap is the largest in the coolest phases (below 5000 K). Severe degeneracies in the initial masses appear. For instance, a star with log(\teff)=3.6 and \lL=4.2 can be reproduced by tracks of stars with initial masses between 7 and 15 \msun. The corresponding stellar ages are thus very different \citep[see also][]{ssc}. \\

The global intrinsic uncertainty within models computed with the same stellar evolution code (0.05 dex at most for a 20 \msun\ model) is much lower than the uncertainty coming from the use of tracks computed with different stellar evolution codes (0.4 dex at maximum for $M=20$ \msun, see Fig.\ \ref{fig_env_all}). The uncertainty is in both cases larger beyond the TAMS, and it appears that no clear consensus exists on the position of lower end of the red giant branch nor on the temperature of the red giant branch itself, making it very difficult to draw trustful conclusions when comparing effective temperatures and luminosities obtained from spectroscopic analysis with predictions of evolutionary tracks for yellow/red supergiants. Age determinations are also very uncertain.

\begin{table*}[t]
\begin{center}
\caption{Properties of the 20 \msun\ model computed with different codes, without rotation, and evaluated at four different evolutionary phases$^1$.} \label{tab_param}
\begin{tabular}{lcccccc}
\hline \hline
Phase    & STERN$^2$ &  Geneva &  FRANEC  &  Padova & MESA & STAREVOL \\ 
\hline 
\rowcolor{Gray} \sf{ X$_c$} = 0.5   &  & & & && \\
 & & & & & & \\
 \teff (K) & 33522 & 33758 & 33709 & 32538 & 33011 & 33365 \\
 \lL  & 4.764  & 4.733 & 4.736 & 4.747 & 4.745 &  4.733\\
 age (Myr) & 3.598 & 3.745 & 4.079 & 4.279 & 4.073 & 3.963\\
 & & & & & & \\
\hline
\rowcolor{Gray} \sf{TAMS$^3$} &  & & & && \\
 & & & & & & \\
 \teff (K) &  24630& 29297 & 28094 &  25972& 28733 & 27902 \\
 \lL  & 5.096   & 5.001& 5.036 & 5.045 & 5.007 & 4.995  \\
 age (Myr) & 8.173 & 7.819 &9.085  & 9.100 & 8.598 &8.535 \\
 & & & & & & \\
\hline
\rowcolor{Gray} \sf{ZAHeMS} &  & & & && \\
 & & & & & & \\
 \teff (K) & 5388 & 24014 & 27382 & 15657 & 23200 & 24259 \\
 \lL  & 4.987  & 5.025 & 5.012 & 5.044 & 5.042 & 5.015 \\
 age (Myr) & 8.185 & 7.828 & 9.088 & 9.111 &  8.610& 8.541\\
 & & & & & & \\
\hline 
\rowcolor{Gray} \sf{TAHeMS$^3$} &  & & & && \\
 & & & & & & \\
 \teff (K) &  -$^*$ & 3753 & 3817  & 3550 & 3702 & 3462 \\
 \lL  &  -$^*$  & 5.050 &  5.206&  5.090& 5.049  &5.027  \\
 age (Myr) & -$^*$  & 8.713  &  9.690& 9.820  & 9.530& 9.748\\
 & & & & & & \\
\hline
\end{tabular}
\tablefoot{1- when the central H mass fraction is 0.5 (X$_c$ = 0.5); at the terminal age main sequence (TAMS); at the zero age helium burning main sequence (ZAHeMS); at the terminal age helium burning main sequence (TAHeMS).\\
2- for the STERN grid, the data available stop at the beginning of the core He burning phase.\\
3- The TAMS and TAHeMS are defined as the points where the mass fraction of H (resp. He) is lower than 10$^{-5}$. }
\end{center}
\end{table*}

\subsection{Rotation}
\label{s_rot}

The effects of rotation on evolutionary tracks of massive stars have been described in details in the literature. We refer to \citet{mm00} and \citet{langer12} for reviews of the main effects. To summarize, the effects of rotation can be broadly described as follows: 

\begin{itemize} 

\item[$\bullet$] geometrical effects: rapid rotation tends to flatten stars, the equatorial radius becoming larger than the polar radius. As a consequence the equatorial gravity is smaller than the polar gravity. According to the Von Zeipel theorem \citet{vz}, the effective temperature at the pole is larger. 

\item[$\bullet$] effects on transport processes: rotation triggers various instabilities and fluid motions in the stellar interior (e.g. meridional circulation, shear instability) which transport angular momentum and chemical species between the stellar core and the surface. Consequently, the surface abundances and internal distribution of species are strongly affected by rotation. 

\item[$\bullet$] effect on mass loss rate: rotation modifies the surface temperature and gravity as described above. Since radiatively driven winds are directly related to these quantities \citep{cak75}, mass loss rates are also affected. \citet{mm00b} showed that on average, they increase whith the ratio of rotational velocity to critical rotational velocity.

\end{itemize}

\noindent A direct consequence of the mixing processes on the evolution of a star is an increase of the duration of the core hydrogen burning phase. Because of mixing, material above the central convective core is brought to the center, leading to a refuelling of the core hence to an extension of the hydrogen core burning phase duration. The effect is qualitatively the same as overshooting. This is somewhat counterbalanced by the larger luminosity of rotating tracks, caused by a different radial profile of the mean molecular weight. A higher luminosity translates into a shorter nuclear burning timescale. But this effect is smaller than the effect of mixing, so that on average, rotation leads to longer nuclear burning sequences (by 10-20\% for the main sequence). Rotating stars thus end their main sequence with larger luminosities than non rotating stars.

In Fig.\ \ref{fig_env_rot} we show the envelopes of tracks for $M=20$ \msun. The dark grey one corresponds to rotating models, the light grey one to non rotating models. The envelopes have been defined from the tracks computed with STERN, FRANEC, the Geneva code, STAREVOL and MESA. We have excluded the Padova tracks since they do not include rotation. The MESA and STAREVOL tracks have been computed assuming an initial surface equatorial velocity of 200 and 233 \kms\ respectively. The STERN and FRANEC tracks have an initial velocity of 300 \kms. The Geneva tracks have been computed for a ratio of initial to critical rotation of about 0.4, corresponding, with their definition of the critical velocity, to about 250 \kms\ at the equator. 

The widening of the main sequence described above is visible in Fig.\ \ref{fig_env_rot}: the dark grey region extends over a wider luminosity range from the ZAMS to the TAMS. Beyond the main sequence, the envelope of the rotating models is wider than that of the non-rotating models immediately after the TAMS, and becomes subsequently narrower at temperatures below $\sim$ 20000 K. In the Hertzsprung gap (5000 $<$ \teff\ $<$ 20000 K), the less luminous of the rotating tracks are overluminous by about 0.1 dex in \lL\ compared to the non rotating tracks. The width of the envelope remains large in the coolest phases: 0.2 dex at $\log \teff\ = 3.7$. Since we are using models with different rotational velocities (between 200 and 300 \kms) the width of the corresponding envelope is most likely affected by this dispersion of velocities and is probably an upper limit. If all models had been computed with the same initial velocity, the spread in luminosity would be slightly smaller.\\

The global uncertainty associated with the choice of a specific grid of stellar evolution models is reduced beyond the ZAMS when considering models including rotation. This is essentially due to the fact that the decrease in luminosity after the TAMS  when the tracks crosses the Hertzsprung gap, is much less important in the rotating models generated with STERN and FRANEC codes. We may attribute it to the larger mass loss and efficient mixing. \\


\section{Comparisons with observational results}
\label{s_comp_obs}

After comparing results of calculation with different codes, we now
turn to comparisons to observational data. From now on, we use only
the publicly available grids of tracks of \citet{brott11a},
\citet{ek12} and \citet{cl13}. All include rotational mixing.

\subsection{The main sequence width}
\label{s_ms}

\begin{figure}[t]
     \centering
          \includegraphics[width=6.5cm,height=5.9cm]{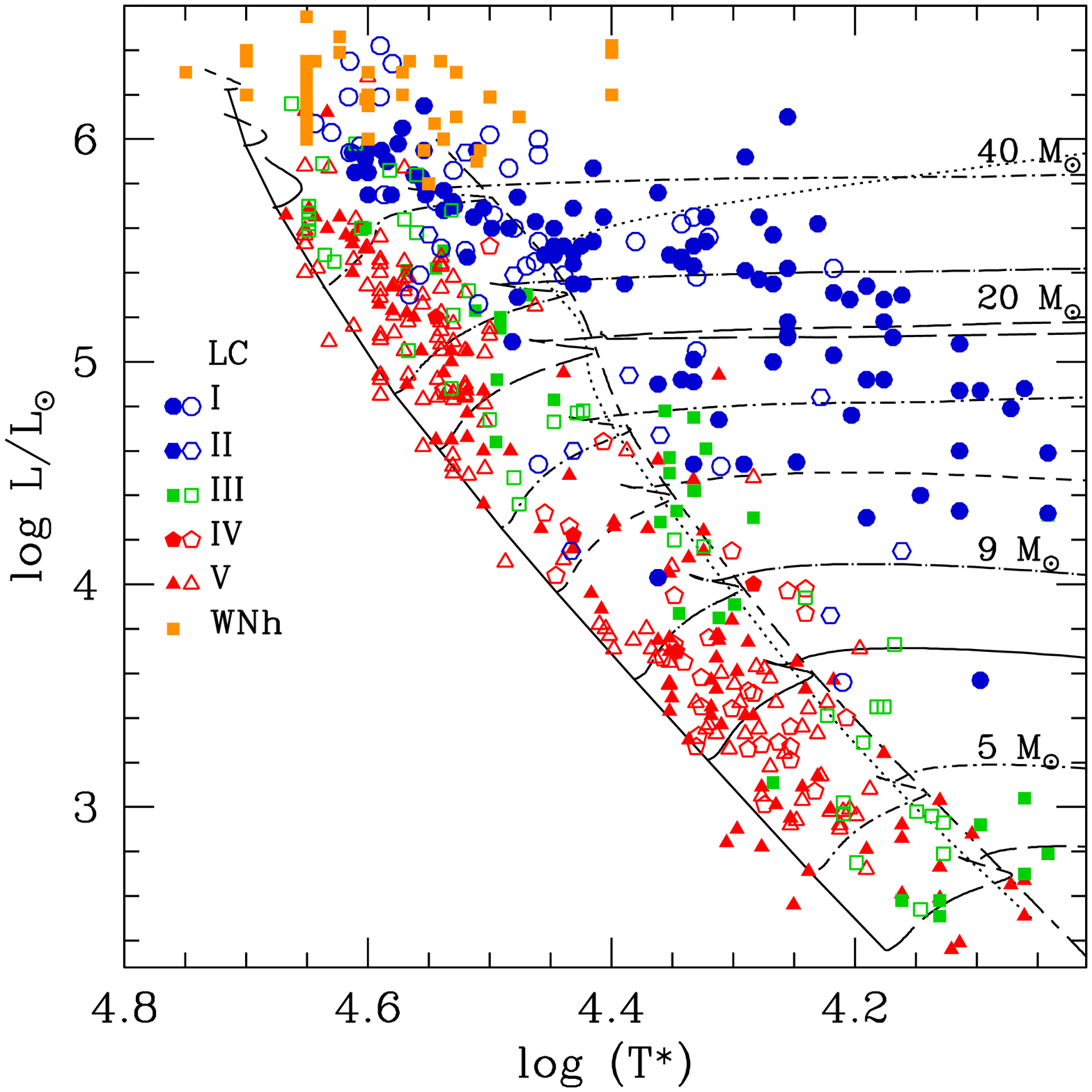}
          \includegraphics[width=6.5cm,height=5.9cm]{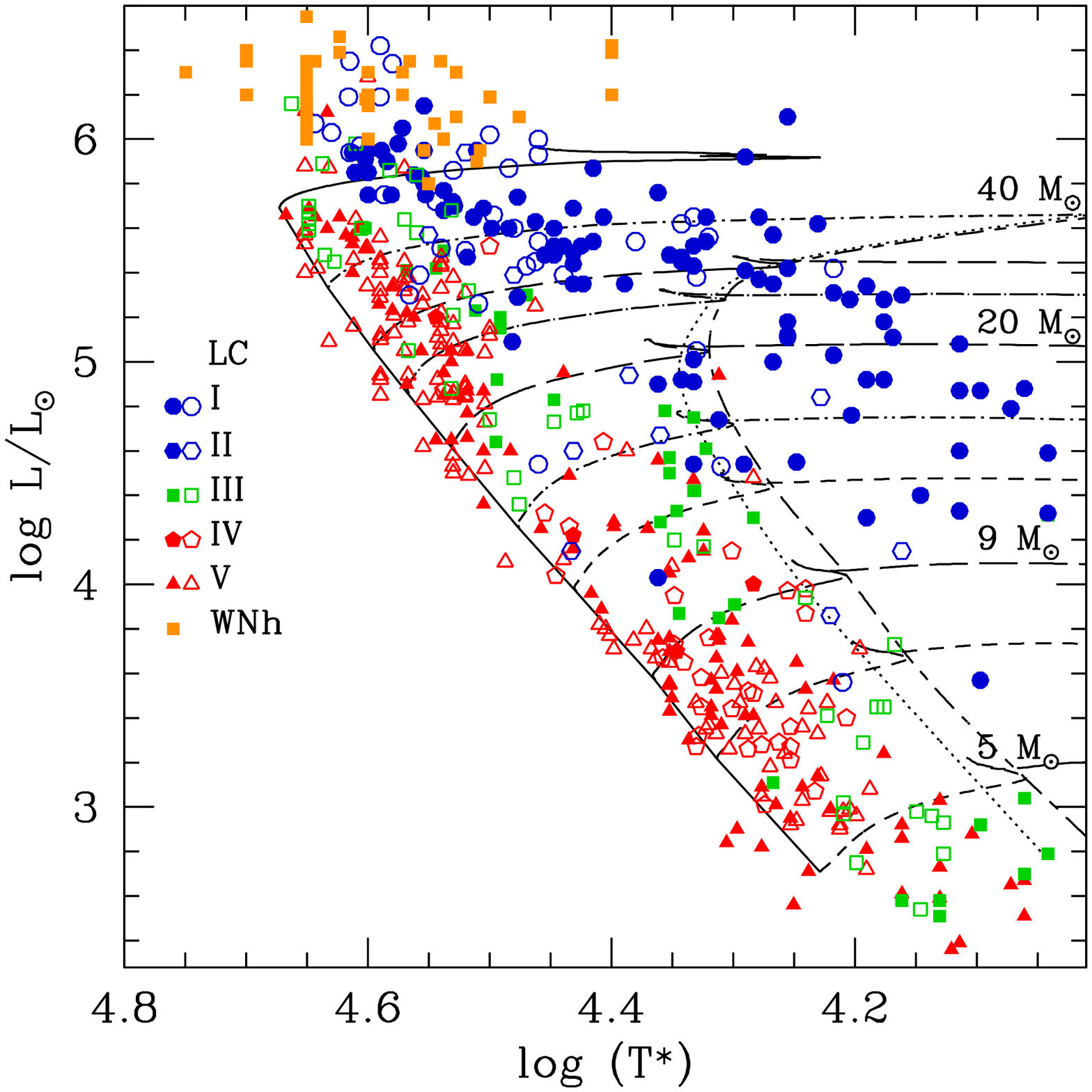}
          \includegraphics[width=6.5cm,height=5.9cm]{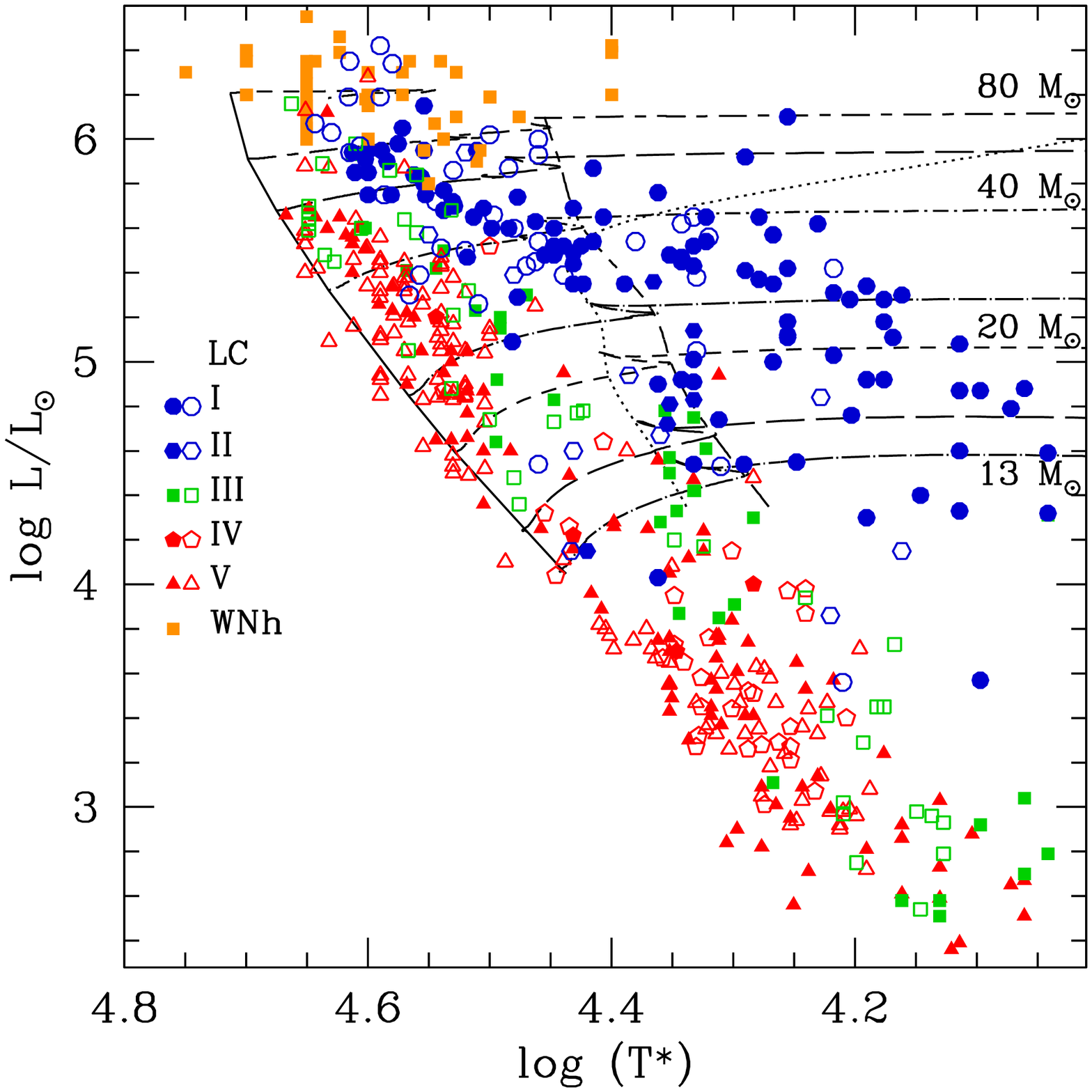}
     \caption{Comparison between evolutionary tracks (black lines) and the location of OB stars in the HR diagram. The evolutionary tracks are from \citet{ek12} (top), \citet{brott11a} (middle) and \citet{cl13} (bottom). They are shown for hydrogen mass fraction lower than 0.60 for clarity. The short dashed - long dashed line connects the cooler edge of the main sequence for the different models and defines the TAMS. The dotted lines is the same for the non-rotating models. Different symbols correspond to different luminosity classes. Filled symbols correspond to stars for which the stellar parameters have been determined through a tailored analysis, while open symbols are for stars with parameters taken from calibrations according to their spectral type. The data sources are listed in the text , \S~\ref{s_ms}.}
     \label{fig_hr}
\end{figure}

In Fig.\ \ref{fig_hr} we show a HR diagram with OB stars and the evolutionary tracks of the three public grids. The tracks are truncated so that only the part corresponding to a hydrogen mass fraction higher than 0.60 is shown\footnote{This value is chosen so that the end of the main sequence for the 40\msun\ track is visible.}. This is already a high value for O stars, corresponding to He/H$=$0.35 by number. The stellar parameters for the comparison stars result from detailed analysis with atmosphere models or from calibrations according to spectral types. In that case, the calibrations of \citet{msh05} have been used to assign an effective temperature. A bolometric correction was subsequently computed following \citet{mp06}. Extinction was calculated from B$-$V. The distance to the stars (taken from parallaxes when available, or from cluster membership) finally lead to the magnitudes and thus, with the bolometric correction, to the luminosity. The data concerning the comparison stars have been taken from the following studies: \citet{mj93}, \citet{mcerlean99}, \citet{vrancken00}, \citet{walborn02}, \citet{lyu02}, \citet{lev04}, \citet{ww05}, \citet{paul06b}, \citet{paul06bsg}, \citet{melena08}, \citet{searle08}, \citet{mark08}, \citet{arches}, \citet{hunter09}, \citet{paul10}, \citet{lef10}, \citet{lier10}, \citet{neg10}, \citet{przy10}, \citet{ngc2244}, \citet{abmag} and \citet{jc12}. In the following, we will assume that the main sequence is populated by stars of luminosity class V and IV. This is a reasonable assumption for stars below $\sim$ 40 \msun\, but certainly an oversimplification for stars above 40 \msun\ for which the spectroscopic luminosity class is not necessarily matching the internal evolutionary status (a luminous star with a strong mass loss can still be burning hydrogen in its core and already have the appearance of a supergiant because of its wind).

The upper panel of Fig.\ \ref{fig_hr} shows the Geneva tracks with initial rotation on the ZAMS between 180 and 270 \kms\ depending on the initial mass. The width of the main sequence (between the solid and short dashed - long dashed lines) corresponds well to the position of main sequence stars below $\sim$ 7 \msun. The extension of the main sequence might be slightly too small for masses between 7 and 25 \msun\ (compared with the location of red triangles and pentagons, that is stars with luminosity classes of IV and V ). The main sequence width is larger when rotation is included, as expected. \\
In the middle panel of Fig.\ \ref{fig_hr} we show the HR diagram built using the tracks of \citet{brott11a} for \vsini\ = 300 \kms. The main sequence for models including rotation is wider than for the Geneva models. For masses of 10--15 \msun, the main sequence extension corresponds to the area populated by luminosity class V, IV and III objects. Bright giants and supergiants are located beyond the main sequence. All giant stars (green squares) being included in the main sequence width, the core H-burning phase in the Brott et al.\ models is too extended. The models without rotation have a narrower main sequence, in better agreement with the position of dwarfs and sub-giants below $\sim$ 15 \msun.  
The main difference between the Geneva and STERN tracks is attributed to the overshooting parameter ($\alpha$=0.335 for STERN versus 0.1 for Geneva). The larger overshooting in the Brott et al. models translates into a wider main sequence. This is particularly true above $\sim$ 30 \msun\ where all blue supergiants are within the main sequence width. Even if some supergiant stars can in principle still be main sequence objects at high luminosity, it is unlikely that all of them are core-H burning objects, indicating that an overshooting of 0.335 is too large for stars with masses above 20 \msun. \\
Finally, the bottom panel of Fig.\ \ref{fig_hr} shows the FRANEC models of \citet{cl13}. Unfortunately, this grid only includes models for masses larger than 13 \msun\ so we focus on the HR diagram above this mass. Between 13 and 20 \msun, the main sequence width appears to be a little wider than the extension of the region where luminosity class V and IV stars are located. Most giants (green squares) are within the predicted main sequence band. Non rotating models better account for the observed extension of the main sequence. Beyond 20 \msun\ and up to 40 \msun\ the MS width remains roughly constant and includes many supergiants. It might thus be too wide. As for the STERN models, this may be due to the overshooting parameter ($\alpha$=0.2 for Chieffi \& Limongi). We note the peculiar behaviour of the non rotating models above 40 \msun: the core H burning phase extends to cooler temperatures than the models including rotation.

From the HR diagram, we thus conclude that moderate values of the overshooting parameter ($\alpha <$ 0.2) in models with initial rotational velocities of 250--300 \kms\ correctly reproduce the main sequence width. For models without rotation, a larger amount of overshooting is required to compensate for the reduction of the main sequence width. 


There are few determinations of the strength of overshooting in the
literature, and they provide conflicting results. \citet{ribas00} used
eclipsing binaries to show that $\alpha$ increases with mass, but
\citet{claret07} found that $\alpha$=0.2 could reproduce correctly the
properties of 3 $<$ M $<$ 30 \msun\ stars. Using asteroseismology,
\citet{briquet07} determined $\alpha$=0.44$\pm$0.07 for the B2IV star
$\theta$ Oph and \citet{briquet10} obtained $\alpha$=0.1$\pm$0.05 for
the O9V star HD~46202. One of the main reasons for these
  differences is that it is usually the extent of the convective core
  that is constrained from observations. From this size, the
  overshooting distance is determined by subtraction of the convective
  core predicted by models in absence of overshooting. This core size
  depends on the input physics and thus varies from model to model,
  implying that the estimates of the overshooting distance also depend
  on the uncertainties of the evolutionary models. Although it relies
  on a physical effect, overshooting can be partly viewed as an
  adjustment variable to correctly reproduce the extension of stellar
  convective cores that are deduced from observations. Keeping this limitation in mind, we can test the values of overshooting adopted in the various codes. \\

\begin{figure}[t]
     \centering
          \includegraphics[width=7cm,height=6.1cm]{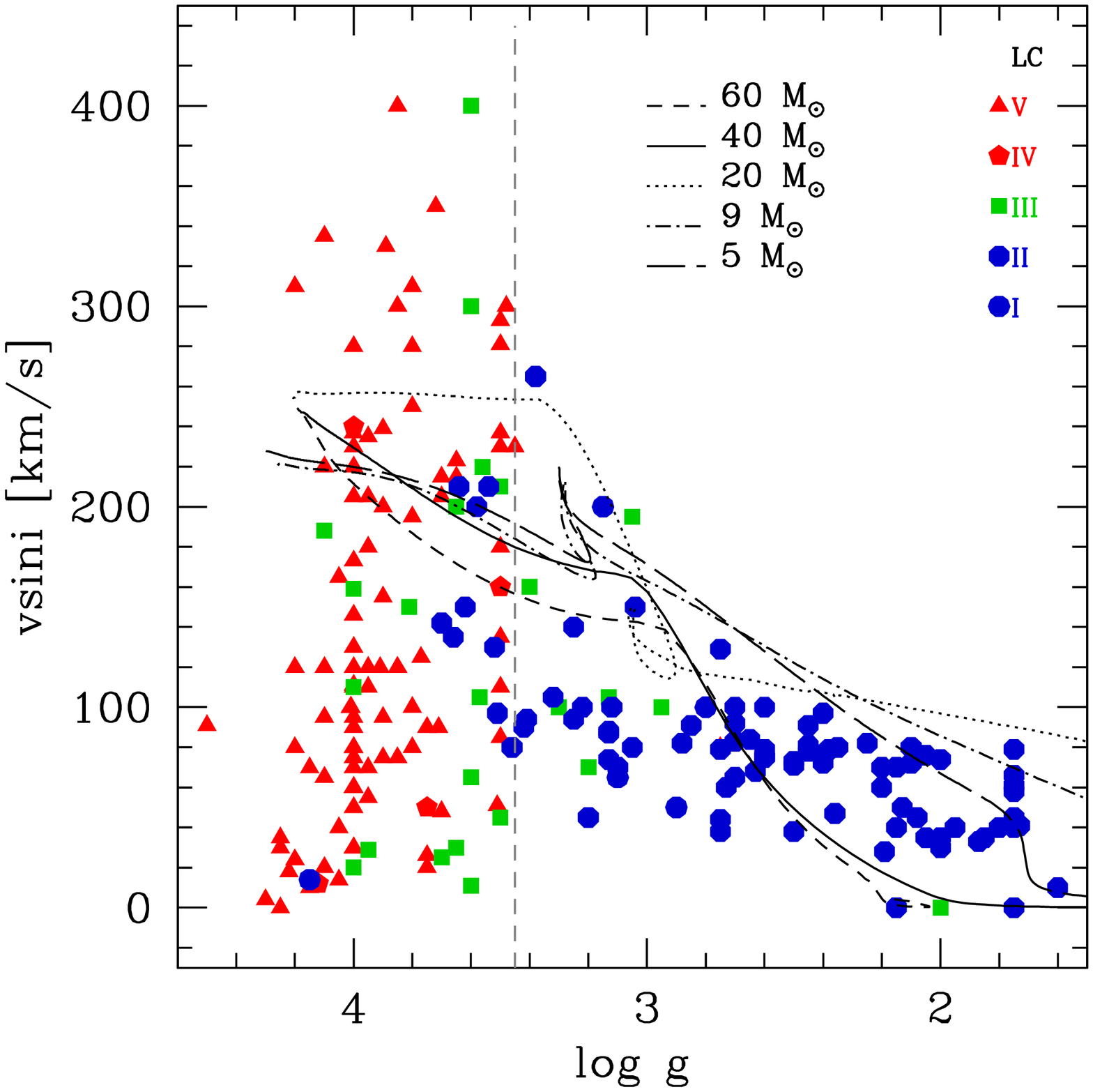}
          \includegraphics[width=7cm,height=6.1cm]{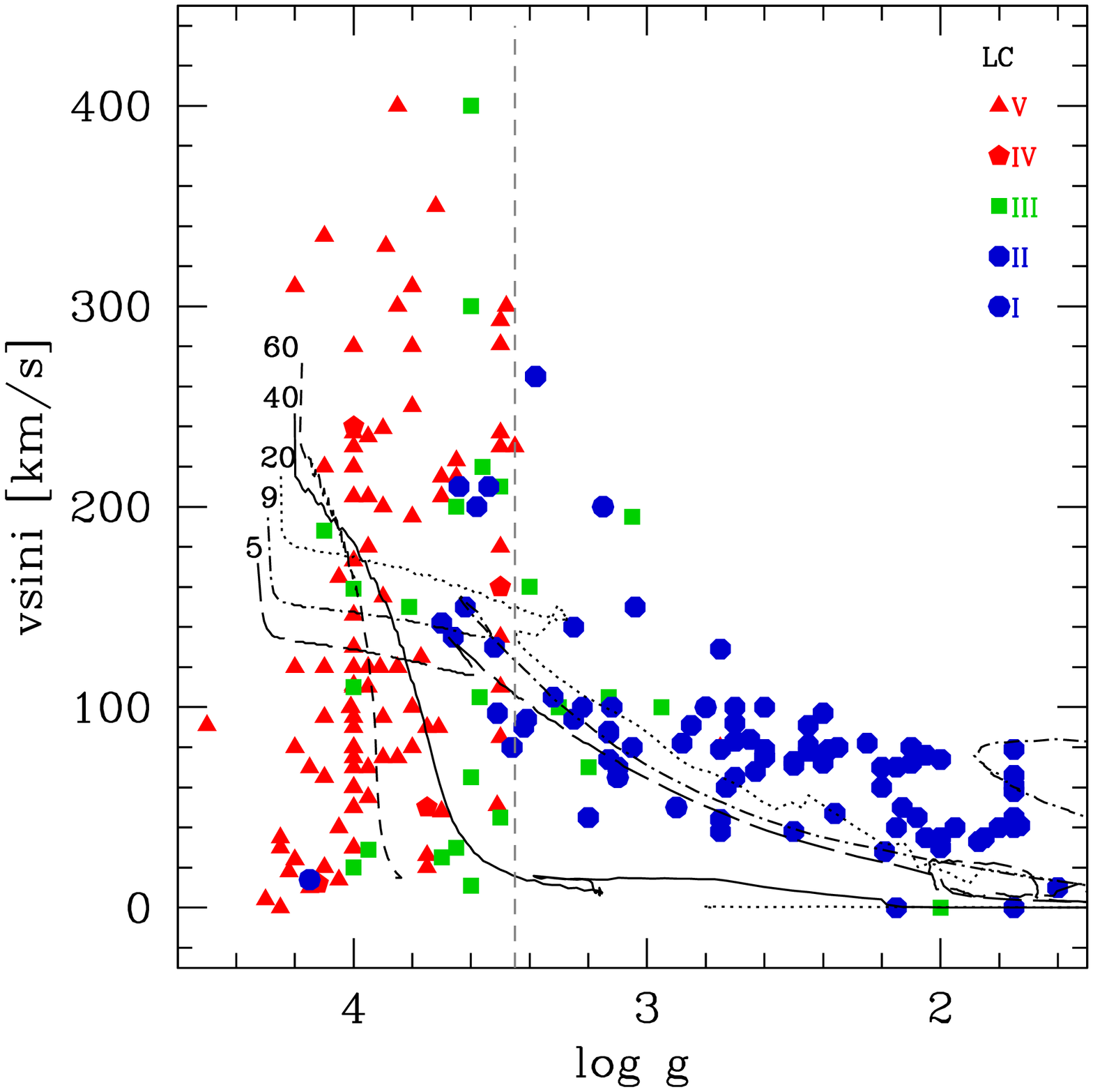}
          \includegraphics[width=7cm,height=6.1cm]{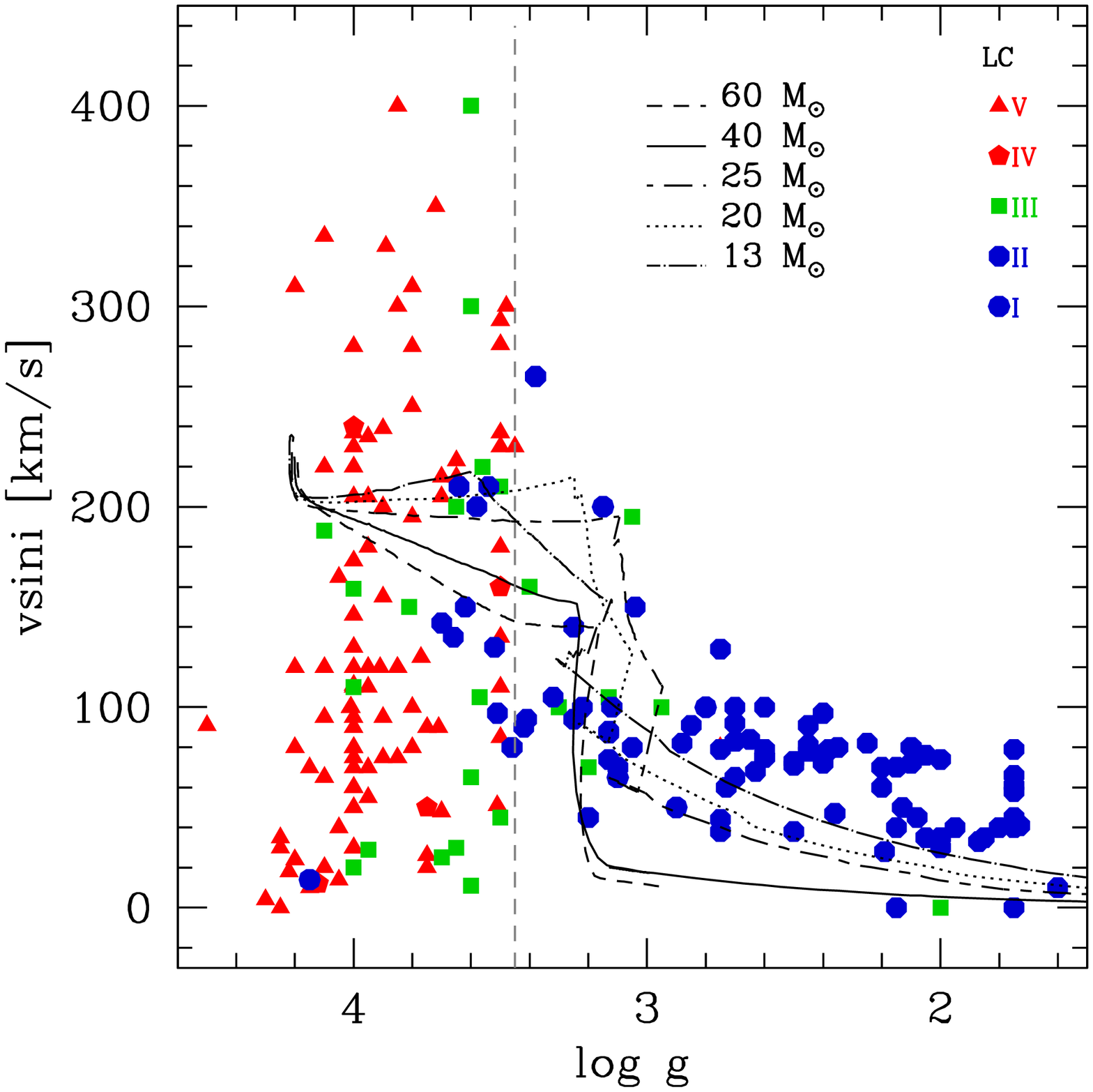}
     \caption{Projected rotational velocity as a function of surface gravity. The symbols have the same meaning as in Fig.\ \ref{fig_hr}. \textit{Top}: evolutionary tracks from \citet{ek12}. \textit{Middle}: evolutionary tracks from \citet{brott11a}. \textit{Bottom}: evolutionary tracks from \citet{cl13}. The vertical dashed line is a guideline located at \logg\ = 3.45.}
     \label{v_logg}
\end{figure}

The large overshooting parameter adopted by \citet{brott11a} results from the comparison of the projected rotational velocity to the surface gravity of massive stars in LMC clusters. High gravity objects show a wide range of rotational velocities, while low \logg\ stars have \vsini\ $<$ 100 \kms. The transition occurs at about \logg\ = 3.2 in the LMC stars and corresponds to the rapid inflation of the star after the end of the core-H burning phase. The larger radius immediately translates into a lower velocity. \citet{brott11a} calibrated the overshooting parameter of their 16 \msun\ model to reproduce this sudden drop. 

Fig.\ \ref{v_logg} shows the \vsini\ -- \logg\ diagram for the stars used to build the HR diagram of Fig.\ \ref{fig_hr}. Only the stars for which both the projected rotational velocity and surface gravity are available are included in Fig.\ \ref{v_logg}. As expected, a clear transition is seen at \logg\ = 3.45. Above this value, velocities span the range 0-400 \kms, while below only the lowest velocities ($<$ 100 \kms) are observed. Main sequence objects (red triangles) are all located on the left part of the transition, while most supergiants (blue circles) are found on the right side. This is qualitatively consistent with the results of \citet{hunter09} and \citet{brott11a}. However, the value we obtain for the transition gravity at solar metallicity is larger than that obtained for the LMC by these authors: 3.45 versus 3.2. The overshooting used by Brott et al.\ to reproduce the properties of LMC stars is too high for Galactic targets. This is illustrated in the upper panel of Fig.\ \ref{v_logg} where we see that the Brott et al's tracks have a terminal main sequence (identified by the loop in the tracks) at \logg\ = 2.9--3.2 for masses between 5 and 20 \msun. The large overshooting increases the core-H burning phase, resulting in a bigger star and thus a lower gravity at the hydrogen core exhaustion. The middle panel of Fig.\ \ref{v_logg} shows the behaviour of the Geneva models. In the 5--20 \msun\ range, the TAMS is located at \logg\ = 3.25--3.60, in better agreement with the observational data. However, the 5 and 9 \msun\ tracks have a TAMS at \logg\ slightly too large compared to the position of luminosity class V stars. A value of $\alpha$ equal to 0.1 is a little too low to reproduce the bulk of main sequence stars. The bottom panel presents the tracks from \citet{cl13} in the mass range 13 \msun\ to 60 \msun. The overshoot parameter used for these models is $d/H_p = 0.2$, and we see an intermediate behaviour compared to the two other families of tracks, as expected for this intermediate value of overshooting. The TAMS is located at \logg\ = 3.0--3.2, at gravities too low compared to the observations. \\ 
Let us note that the Geneva and FRANEC tracks (Fig~\ref{v_logg}b and c) behave similarly beyond the TAMS (in particular the 40 \msun\ track) and only cover the slowest supergiants (blue circles), with predicted surface velocities of less than 40 \kms for log $g$ $<$ 3.0. 
The models from the STERN grid on the other hand experience a weaker braking and conserve a more rapid rotation beyond. They better reproduce the observed distribution of rotation velocities of supergiants. The treatment of rotation induced mixing and angular momentum evolution in the STERN code is different from that implemented in the Geneva and FRANEC codes, which might explain these differences in the evolution of the surface equatorial velocity.\\

In conclusion, an overshooting parameter a little larger than 0.1 is required to have a convective core extension able to account for the observed width of the main sequence for Galactic OB stars.

\begin{figure}[]
\centering
\includegraphics[width=9cm]{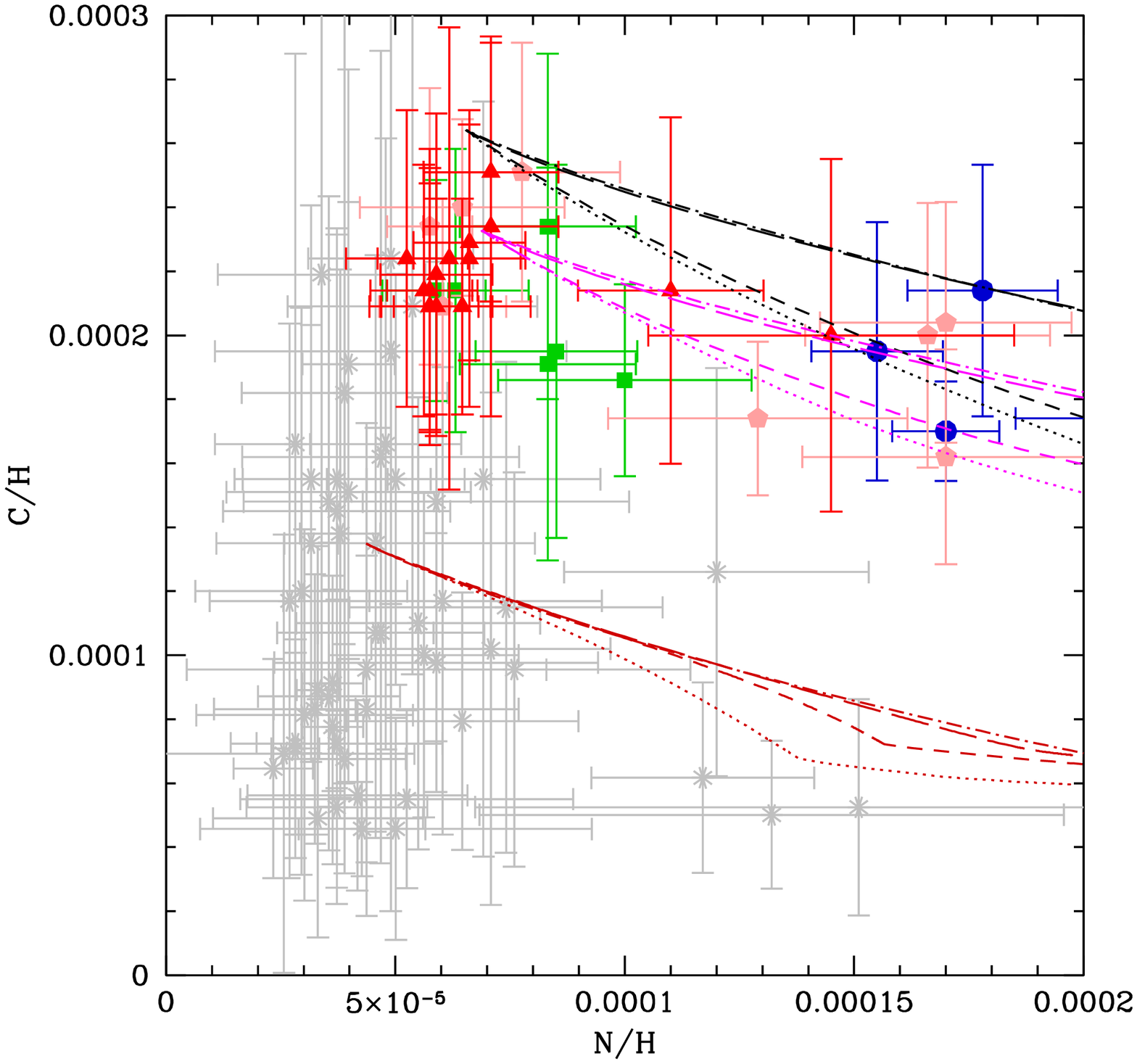}
\caption{C/H as a function of N/H. The symbols correspond to results from spectroscopic analysis from \citet{arches,hunter09,przy10,nieva11,firnstein12,jc12}. The evolutionary tracks for various masses between 15 and 60 \msun\ are taken from \citet{ek12} (black), \citet{brott11a} (red) and \citet{cl13} (magenta).}\label{fig_CN_zoom}
\end{figure}

\subsection{Surface carbon and nitrogen content}
\label{s_ab}

\begin{figure}[t]
\centering
\includegraphics[width=9cm]{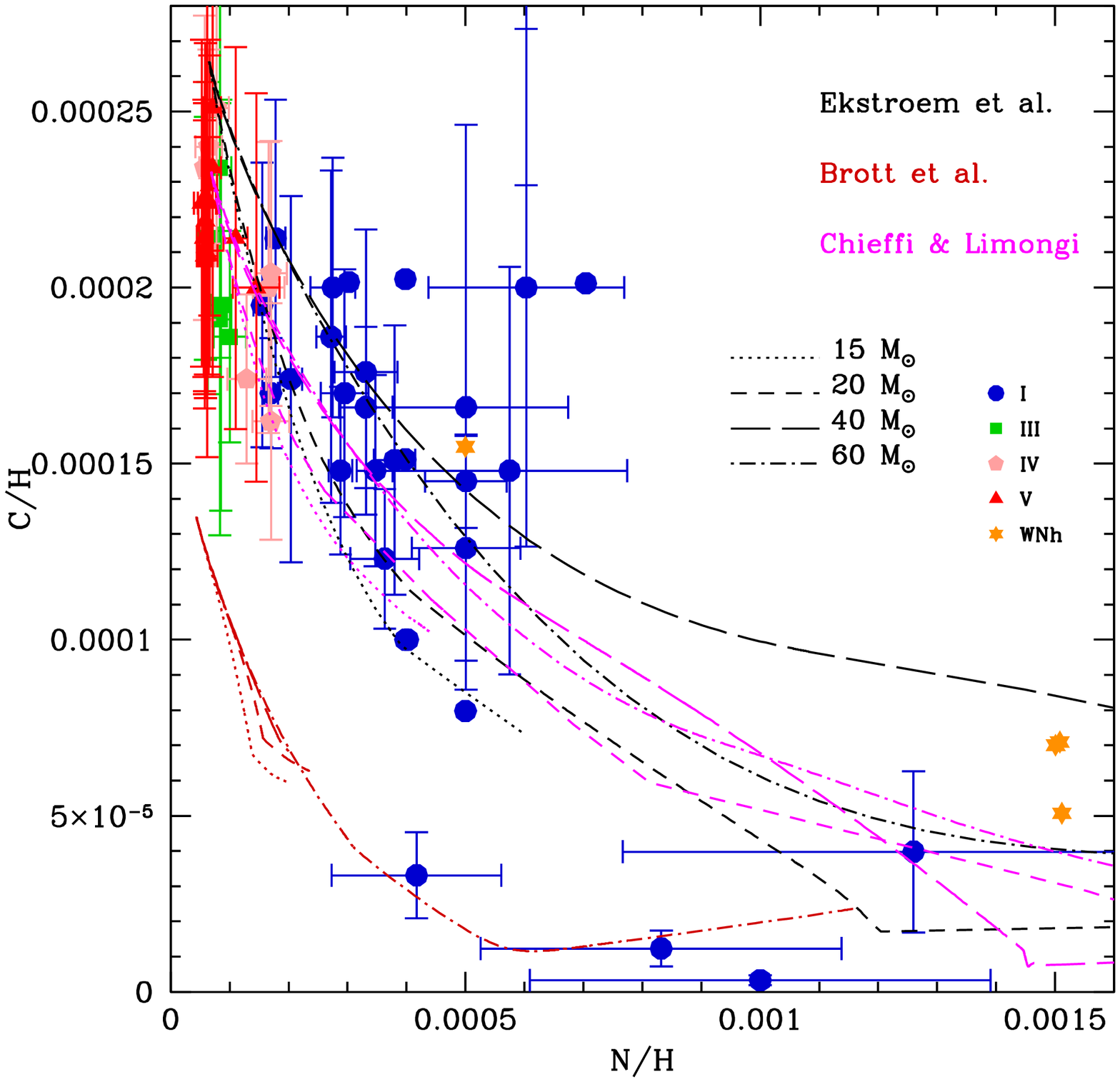}
\caption{Same as Fig.\ \ref{fig_CN_zoom} on a wider scale to show abundances of evolved objects. The stars analyzed by \citet{hunter09} have been omitted for clarity.}\label{fig_CN}
\end{figure}

Rotation modifies the surface abundances because the of the induced mixing processes. Surface abundances are thus important diagnostics to see how realistic is the treatment of rotation in evolutionary models.
In Fig.\ \ref{fig_CN_zoom} and \ref{fig_CN} we compare the C/H and N/H ratios predicted by the models to those determined for OB stars. The symbols correspond to Galactic stars analyzed by means of quantitative spectroscopic analysis. The evolutionary tracks of \citet{brott11a}, \citet{ek12} and \citet{cl13} are shown. They span a mass range between 15 and 60 \msun. OB stars on the main sequence have an initial carbon content between 5.0$\times 10^{-5}$ and 2.6$\times 10^{-4}$. We note that values lower than 2.0$\times10^{-4}$ are only obtained in the study of \citet{hunter09}. The other studies of main sequence stars \citep{przy10,nieva11,firnstein12} all find C/H larger than 2.0$\times 10^{-4}$, consistent with the solar reference value of \citet{ga10}. 

The \citet{brott11a} tracks have an initial composition roughly corresponding to the average of the B stars of \citet{hunter09}. The few B stars of Hunter et al.\ with evidence for N enrichment are rather well accounted for by these tracks. On the other hand, the OB stars analyzed by \citet{przy10}, \citet{nieva11} and \citet{firnstein12} cannot be reproduced by the Brott et al.\ tracks. The \citet{ek12} and \citet{cl13} predictions are better (although on average slightly too carbon rich for the Geneva models) to account for the initial composition and the evolution of these stars. One may wonder whether the samples of \citet{hunter09} and \citet{przy10}/\citet{nieva11}/\citet{firnstein12} are taken from environments with different metallicities. \citet{hunter09} focused on NGC 6611 towards the Galactic Center, and NGC3293 / NGC 4755 at Galactic longitudes of 285$^{o}$ and 303$^{o}$ and distances close to 2 kpc. The stars of \citet{przy10}/\citet{nieva11}/\citet{firnstein12} are mostly from the solar neighborhood. We thus do not expect strong metallicity differences among the stars shown in Fig.\ \ref{fig_CN_zoom}. The reason for the differences in C abundance at the beginning of the evolution between the two sets of stars is thus not clear. It might be related to the methods used to determine the stellar parameters and the surface abundances. 

Turning to Fig.\ \ref{fig_CN}, we see that evolved OB stars are well reproduced by the Ekstr{\"o}m et al.\ and Chieffi \& Limongi models, although the number of stars with good C and N abundances is still too small to draw firm conclusions. The \citet{brott11a} tracks have a too low carbon content at a given N/H because of the initial offset discussed above. 

These comparisons indicate that accurate abundance determinations are necessary before the predictions of evolutionary models regarding surface composition can be tested at depth. The reason for the difference between the study of Hunter et al.\ on the one hand, and of Nieva, Przybilla, Firnstein and collaborators on the other hand needs to be understood before preference can be given to a set of tracks.

\subsection{The upper HR diagram}
\label{hr_up}

\begin{figure*}[t]
     \centering
     \subfigure[Geneva - with rotation]{
          \includegraphics[width=.45\textwidth]{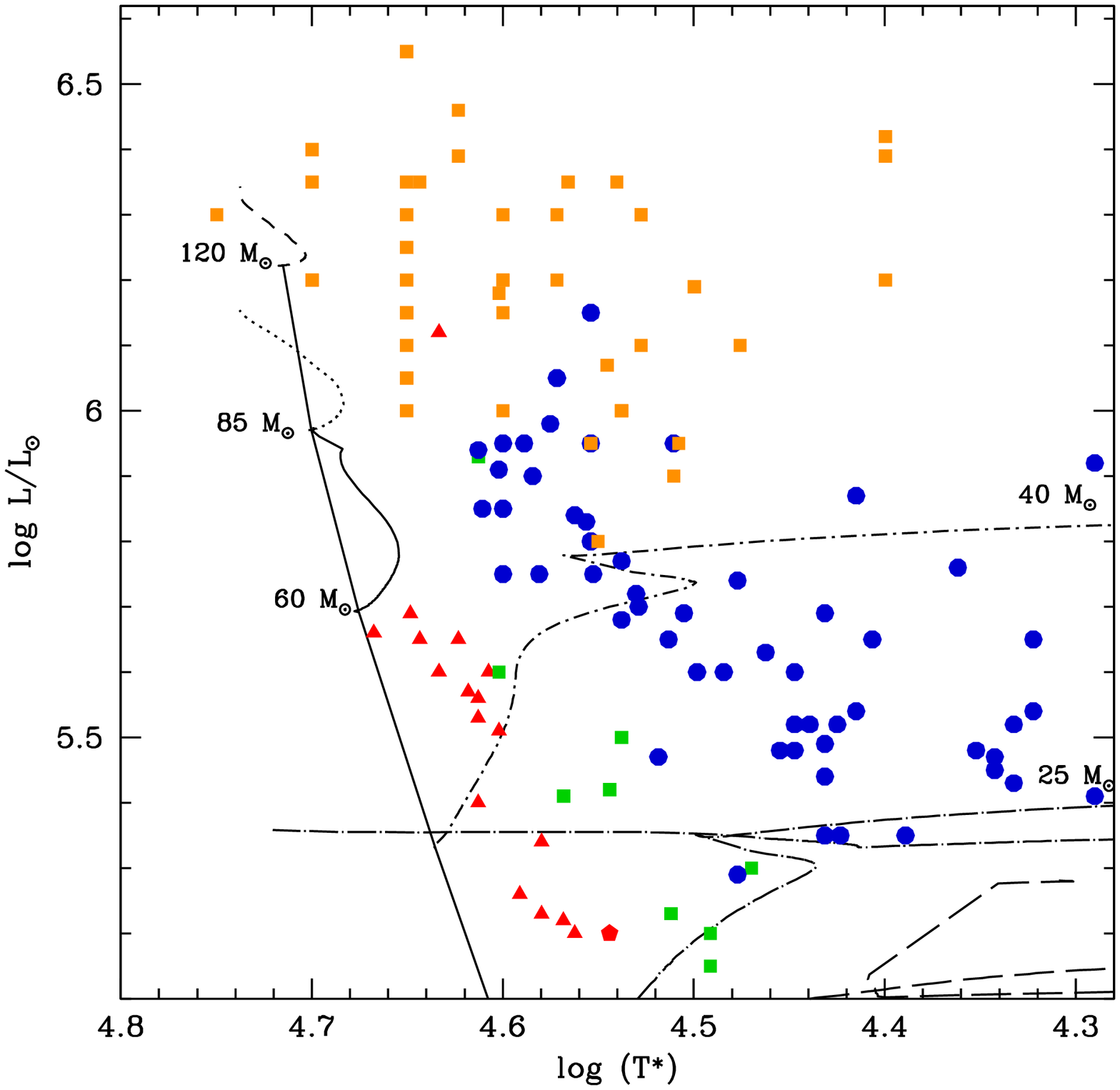}}
     \hspace{0.2cm}
     \subfigure[Geneva - without rotation]{
          \includegraphics[width=.45\textwidth]{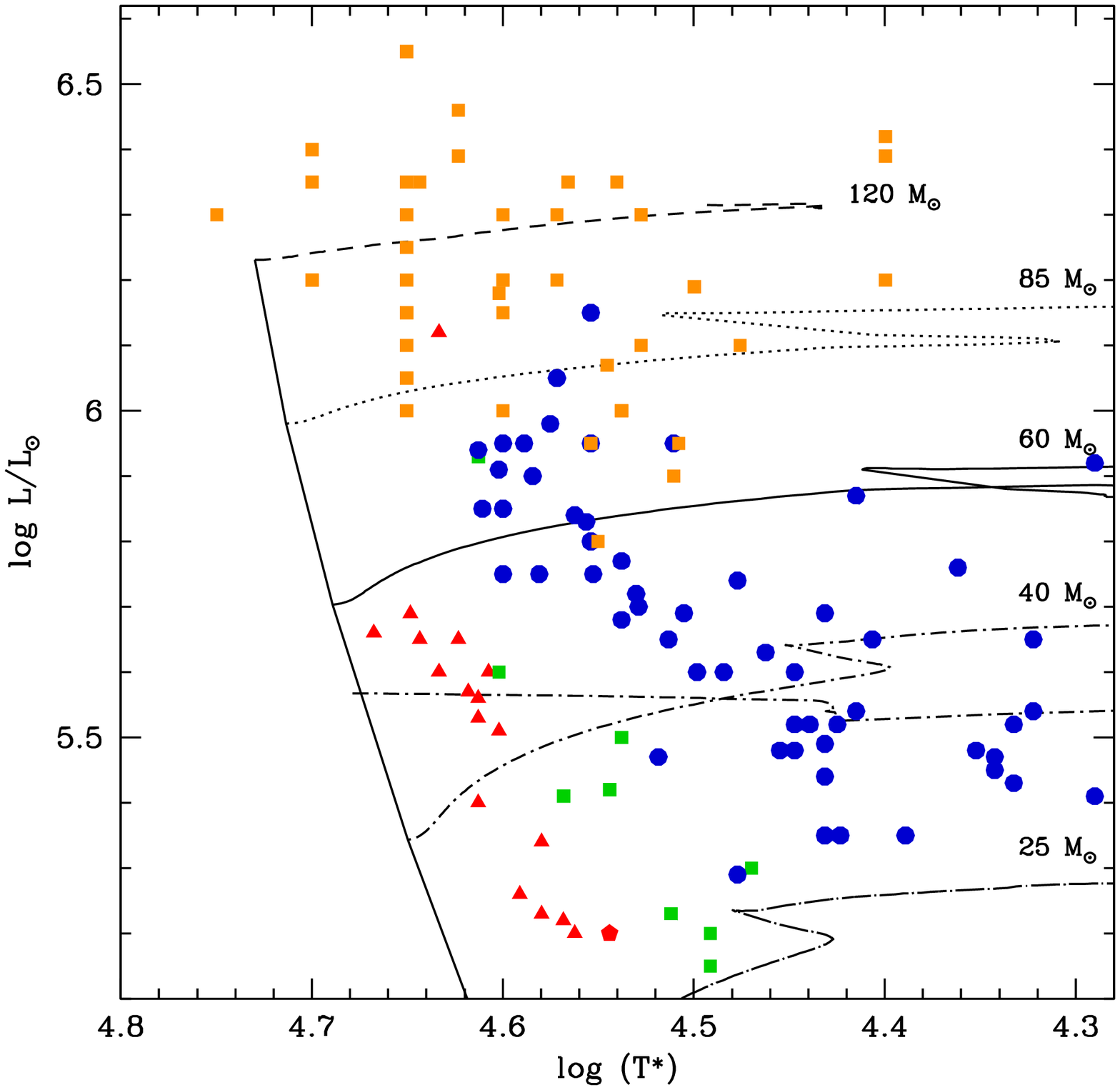}}
     \subfigure[FRANEC - with rotation]{
          \includegraphics[width=.45\textwidth]{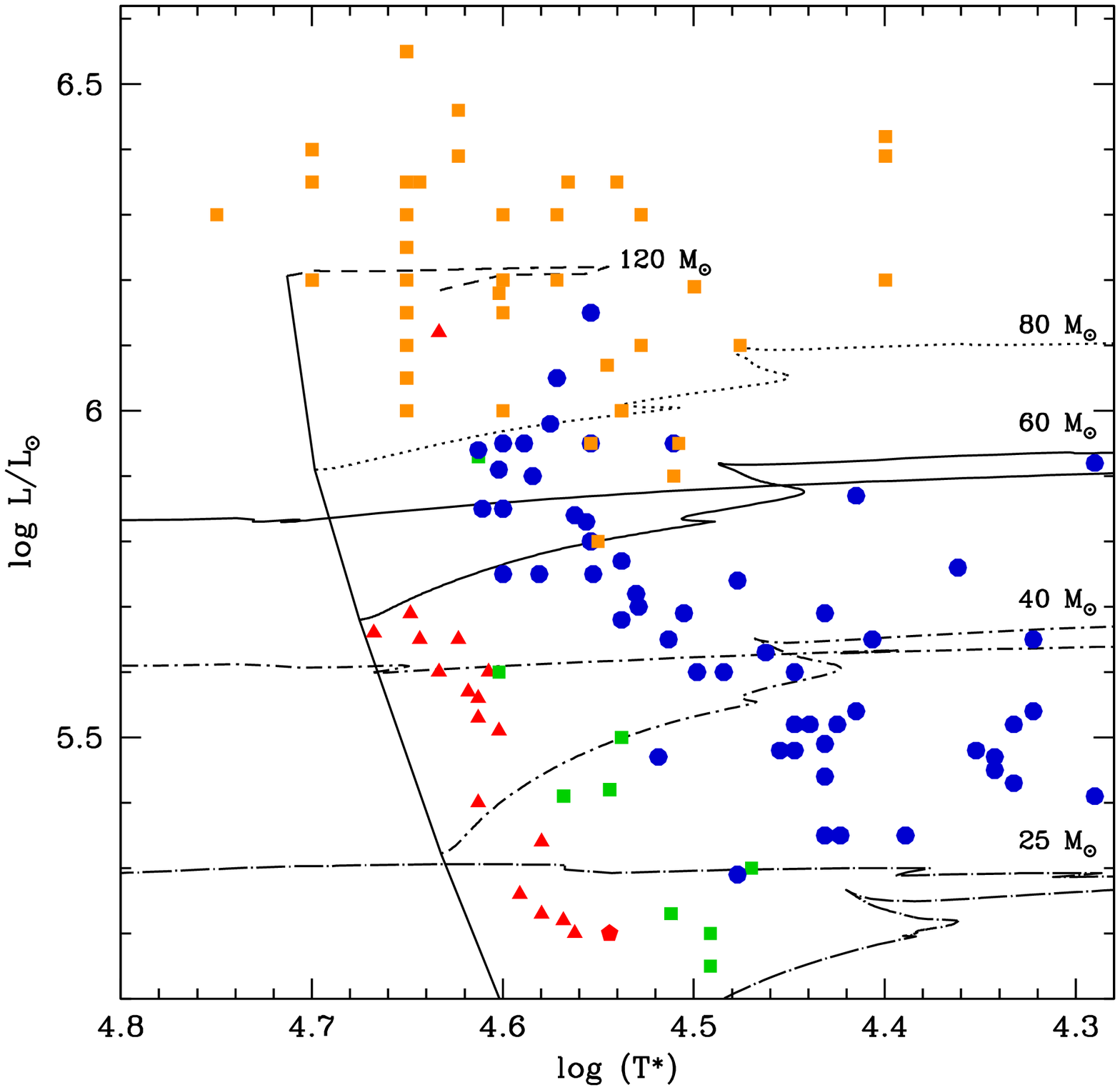}}
     \hspace{0.2cm}
     \subfigure[FRANEC - without rotation]{
          \includegraphics[width=.45\textwidth]{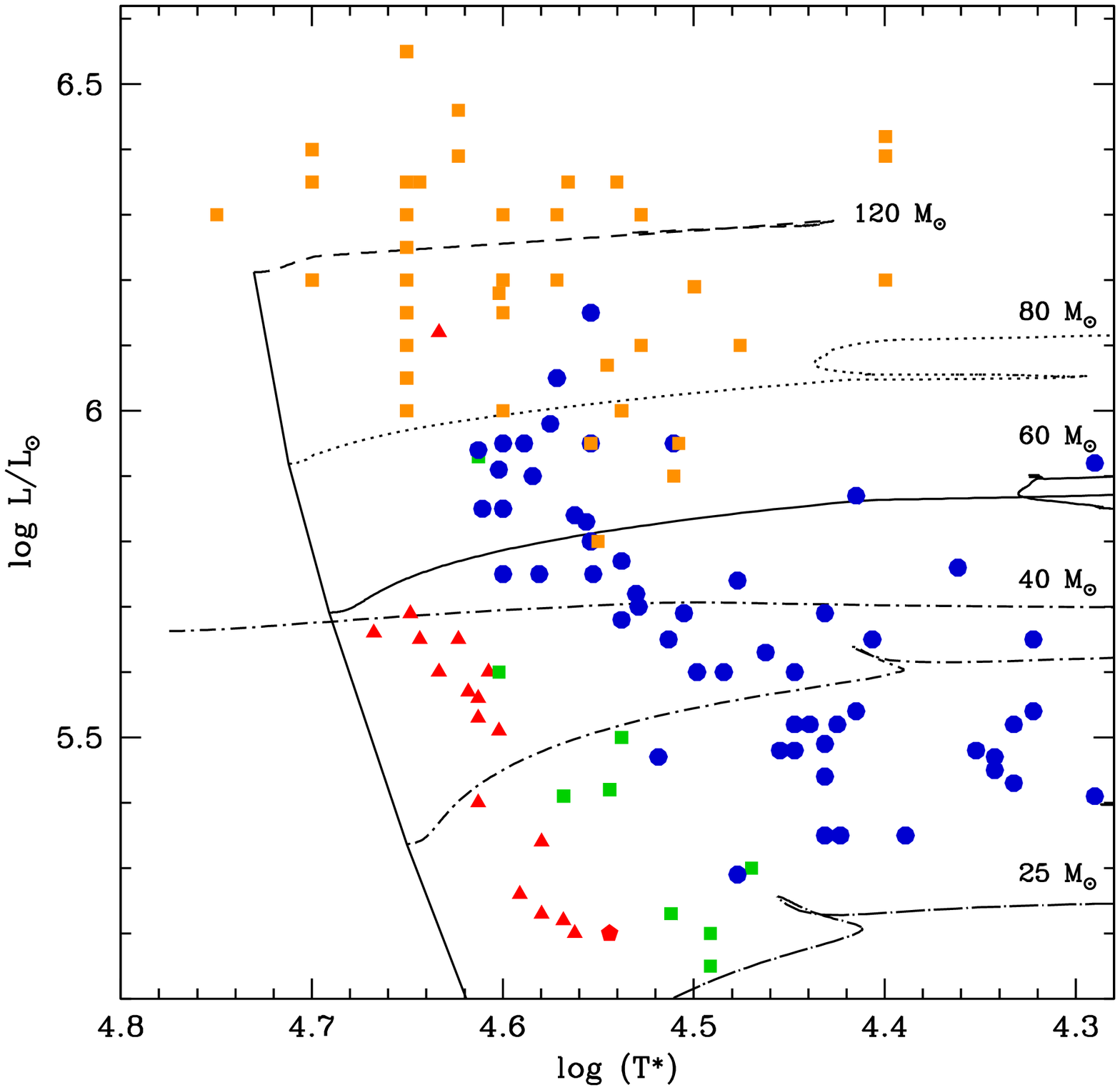}}
     \caption{Same as Fig.\ \ref{fig_hr} but focusing on the upper part of the HR diagram. The orange squares correspond to WNh stars. We have used stars for which stellar parameters determined from spectroscopic analysis are available. In the upper (lower) panels the Geneva (FRANEC) models are shown. The tracks have been plotted for X(He)$<$0.75 which corresponds to He/H$\sim$0.8 by number. \textit{Left}: tracks with rotation. \textit{Right}: non rotating tracks.}
     \label{fig_hr11_up}
\end{figure*}

In this section, we focus on the upper HR diagram, i.e. at luminosities larger than $3 \times\ 10^{5}$ L$_{\odot}$. The only publicly available tracks including rotation for stars with masses larger than 60 \msun\ are provided by the Geneva group and \citet{cl13}. In Fig.\ \ref{fig_hr11_up} we plot the evolutionary tracks of \citet{ek12} and Chieffi \& Limongi in the upper and lower panels respectively. Only the part of the tracks with a surface helium mass fraction smaller than 0.75 is shown. This is equivalent to a number ratio He/H of about 0.8. This value is the largest determined for most WNh stars \citep{hamann06,arches,lier10}. 

We first focus on the Geneva models. The WNh stars and the most luminous supergiants cannot be accounted for by the evolutionary tracks with rotation of \citet{ek12}. The 60, 85 and 120 \msun\ tracks turn to the blue almost immediately after the zero age main sequence (Fig.\ \ref{fig_hr11_up}, upper left panel). The 60~\msun\ track later comes back to the red part of the HRD, but at that time it is depleted in hydrogen, while the comparison stars all still contain a significant fraction of hydrogen. The 85 and 120 \msun\ tracks do not come back to the right of the ZAMS and cannot reproduce the position of the WNh stars. 
We thus conclude that the solar metallicity tracks with rotation of \citet{ek12} do not account for the most massive stars and thus should not be used for comparisons with normal stars with masses above 60 \msun.  
The Geneva tracks are computed for $\frac{v_{eq}}{v_{crit}} \sim 0.4$ where $v_{eq}$ and $v_{crit}$ are the equatorial velocity and the break-up velocity. For stars with M $>$ 60 \msun\ the initial velocity is $\sim$280 \kms. Although not extreme, and in spite of the rapid drop of the rotational velocity with age \citep[see Fig.\ 10 of][]{ek12} the fast initial rotation may explain the blueward evolution of the most massive models. This behaviour is similar to that of quasi-chemically homogeneous tracks \citep{maeder87,langer92}.\\
For comparison, we show on the upper right panel of Fig.\ \ref{fig_hr11_up} the Geneva non rotating tracks, with the same cut in helium mass fraction. This time, the 60, 85 and 120 \msun\ models cross the region occupied by the most luminous supergiants and the WNh stars. With a low or moderate helium content, they can thus reproduce the existence of these very luminous objects. One can conclude that the current Geneva rotating tracks are probably computed with a too high initial rotational velocity to explain the existence of most of the Galactic WNh and early O supergiants.
The tracks by \citet{mm03} include rotation with an initial equatorial velocity of 300 \kms\ and have $Z=0.020$. They are super-solar according to the current calibration of solar abundances ($Z=0.014$). They do not show the rapid blueward evolution just after the ZAMS seen in the Ekstr{\"o}m et al.\ tracks. They better account for the very massive stars, as exemplified by the studies of \citet{hamann06,arches}. Since the initial rotational velocity in the \citet{mm03} tracks is larger than in the \citet{ek12} models, we attribute the different behaviour to the metallicity difference. 

\citet{levesque12} used the evolutionary tracks including rotation of \citet{ek12} to compute population synthesis models. They concluded that the inclusion of rotation lead to a significant increase of the ionizing flux in young stellar population. The main reason is a shift towards higher luminosities of the tracks when rotation is included (see Sect.\ \ref{s_rot}). In the earliest phases, the presence of stars with masses in excess of 60 \msun\ also contributes to the larger ionizing flux, since as illustrated above, the tracks predict much higher temperature when rotation is included. As discussed by \citet{levesque12}, their predictions are probably biased towards populations with larger than average rotational velocities. In view of the direct comparison between evolutionary tracks and position of Galactic massive stars in the HR diagram, we confirm that the \citet{ek12} tracks should be used with a good recognition that they are relevant for fast rotating objects.

Coming back to the lower panels of Fig.\ \ref{fig_hr11_up}, we see that the models of Chieffi \& Limongi do not suffer from the same problems as the Geneva models. The 60 and 80 \msun\ tracks including rotation evolve classically towards the red part of the HR diagram. They all have initial rotation of about 250 \kms\, similar to that of the 20\msun\ and 40 \msun\ models from \citet{ek12} as can be see from Fig.~\ref{v_logg}. They can account for stars as luminous as 10$^6$ L$_{\odot}$. The 120 \msun track also evolves redwards, but turns back to the blue at $\log \teff\ = 4.54$. Many of the luminous WNh stars can be explained by this track. Only three objects at $\log \teff\ = 4.4$ are not reproduced by the FRANEC tracks. But the 120 \msun\ non-rotating track (lower right panel of Fig.\ \ref{fig_hr11_up}) extends down to such temperatures. Since the rotation of massive stars spans a range of values, it is possible that some objects are slow rotators. Unfortunately, it is difficult to measure rotation velocities in WNh stars since their spectrum is dominated by lines formed above the photosphere, in the wind.\\
We thus conclude that overall the Chieffi \& Limongi tracks better reproduce the population of luminous O supergiants and WNh stars in the HR diagram. For the most luminous stars, it appears that non-rotating tracks may give a better fit to the observations than rotating ones when only considering the position in the HR diagram. The difference between the FRANEC and Geneva computations are: 1) a larger overshooting and mixing length parameter and 2) the inclusion of efficiency factors in the treatment of rotation in the study of Chieffi \& Limongi. The mass loss rates in the temperature range of interest are the same. Since the Geneva models are very different with and without rotation, we tend to attribute the differences with the FRANEC models to the second point listed above. The efficiency factors of Chieffi \& Limongi reduce the impact of rotation on the diffusion coefficients. Qualitatively, they could limit the strong effects of rotation seen in the grid of \citet{ek12}.



\section{Conclusion}
\label{s_conc}

We have performed a comparison of evolutionary models for massive stars in the Galaxy. The published grids of \citet{bert09}, \citet{brott11a}, \citet{ek12} and \citet{cl13} have been used. We have also computed additional models with the codes STAREVOL \citep{dmp09} and MESA \citep{pax13}. Our goal was to estimate and highlight the uncertainties in the output of these models. Our conclusions are:

\begin{itemize}

\item[$\bullet$] Evolutionary tracks in the HR diagram are sensitive to the adopted solar composition mixture, metallicity, the amount of overshooting and the mass loss rate. The extension of the convective core, due to overshooting and/or rotation-induced mixing, and the adopted initial metallicity are the major sources of uncertainty on the determined luminosity on the main sequence and during the core He burning phase. Within one set of stellar evolution models, we estimate a global intrinsic uncertainty on the luminosity of about $\pm 0.05$ dex for a star with an initial mass of 20 \msun. This uncertainty is lower when studying main sequence stars. This uncertainty translates into an error of about 6$\%$ on the distance, and may rise up to an error of 30$\%$ on the distance of lower mass red supergiants (typically stars with initial mass of 7 \msun).\\

\item[$\bullet$] The evolutionary tracks computed with the six different codes agree reasonably well for the main sequence evolution. Beyond that, large differences appear. They are the largest at low effective temperature. In the red part of the HR diagram, evolutionary tracks with different initial masses and computed with different codes can overlap. The difference on the luminosity of a star located near the red giant branch can be as large as 0.4 dex depending on the stellar evolution models adopted, which makes the estimate of ages and initial masses for red supergiants extremely uncertain.\\

\item[$\bullet$] Comparison of the tracks of \citet{brott11a}, \citet{ek12} and \citet{cl13} with the properties (\teff, luminosity) of a large number of OB and WNh stars in the Galaxy indicate that in the mass range 7--20 \msun\ the Ekstr{\"o}m et al. tracks have a slightly too narrow main sequence width, while the Brott et al. and Chieffi \& Limongi tracks have a too wide one. This is due to the overshooting parameters used. These results are confirmed by the analysis of the distribution of rotational velocities as a function of \logg\ for Galactic stars with various spectral types. A clear drop in rotational velocities is observed at \logg\ = 3.45, at larger gravity than that observed for LMC stars studied by \cite{hunter08}. An overshooting parameter slightly above 0.1 is required to reproduce this trend.\\

\item[$\bullet$] Measurements of surface abundances of carbon and nitrogen are currently too uncertain to help constrain evolutionary models.\\ 

\item[$\bullet$] Stars with initial masses higher than about 60 \msun\ are not accounted for by the Ekstr{\"o}m et al. tracks with rotation. They bend to the blue part of the HR diagram quickly after leaving the zero age main sequence, and do not reproduce the position of the WNh and luminous blue supergiants. Models provided by Chieffi \& Limongi give a much better fit to the observations. Non rotating tracks of both grids can reproduce the position of the most luminous objects.

\end{itemize}

Future analysis of surface abundances of large samples of Galactic OB stars will provide critical constraints on evolutionary models and the treatment of rotation. Better prescriptions of mass loss rates for a wide variety of stars are also required to improve the predictions of evolutionary models.

\begin{acknowledgements}
We warmly thank the MESA team for making their code freely available (http://mesa.sourceforge.net/). We thank Marco Limongi and Alessandro Chieffi for kindly providing their models and associated information. We aknowledge the comments and suggestions of an anonymous referee.
We thank Georges Meynet, Daniel Schaerer, Sylvia Ekstr{\"o}m, Cyril Georgy, Selma de Mink, Hugues Sana for interesting discussions. This study was supported by the grant ANR-11-JS56-0007 (Agence Nationale de la Recherche). 
\end{acknowledgements}

\bibliographystyle{aa}
\bibliography{comp_evolmod.bib}

\end{document}